\documentclass[useAMS]{mn2e}
\usepackage{epsfig}
\def\lesssim{\,\lower2truept\hbox{${<\atop\hbox{\raise4truept\hbox{$\sim$}}}$}\,}
\def\gtrsim{\,\lower2truept\hbox{${>\atop\hbox{\raise4truept\hbox{$\sim$}}}$}\,}

\title[The AT20G extragalactic sample]{The Australia Telescope 20\,GHz (AT20G) Survey: analysis of the extragalactic source sample}
\author[Massardi et al.]{
\parbox[t]{\textwidth}
{Marcella Massardi$^{1}$\thanks{E-mail: massardi@oapd.inaf.it}, Ronald D.\ Ekers$^2$, Tara Murphy$^{3,4}$, Elizabeth Mahony$^{2,3}$, Paul J.\ Hancock$^3$, Rajan Chhetri$^{2,5}$, Gianfranco De Zotti$^{1,12}$, Elaine M.\ Sadler$^3$, Sarah Burke-Spolaor$^{2,6}$, Mark Calabretta$^2$, Philip G.\ Edwards$^2$, Jennifer A. Ekers$^2$, Carole A.\ Jackson$^2$,
Michael J.\ Kesteven$^2$,  Katherine Newton-McGee$^{2,3}$, Chris Phillips$^2$, Roberto Ricci$^7$, Paul Roberts$^2$, Robert J.\ Sault$^8$, Lister Staveley-Smith$^9$, Ravi Subrahmanyan$^{10}$, Mark A.\ Walker$^{11}$, and Warwick E.\ Wilson$^2$}
\vspace*{8pt} \\
$^{1}$INAF, Osservatorio Astronomico di Padova, Vicolo dell'Osservatorio 5, I-35122 Padova, Italy\\
$^{2}$Australia Telescope National Facility, CSIRO Astronomy and Space Science, PO Box 76, Epping, NSW 1710, Australia\\
$^{3}$Sydney Institute for Astronomy, School of Physics, University of Sydney, NSW 2006, Australia\\
$^{4}$School of Information Technologies, University of Sydney, NSW 2006, Australia\\
$^{5}$School of Physics, The University of New South Wales, NSW 2052, Australia\\
$^{6}$Swinburne University of Technology, P.O.\ Box 218, Hawthorn, Vic 3122, Australia\\
$^{7}$INAF, Istituto di Radioastronomia, via Gobetti 101, I-40129 Bologna, Italy\\
$^{8}$School of Physics, The University of Melbourne, Victoria 3010, Australia\\
$^{9}$School of Physics, University of Western Australia, 35 Stirling Highway Crawley, WA 6009, Australia\\
$^{10}$Raman Research Institute, Sadashivanagar, Bangalore 560080, India\\
$^{11}$Manly Astrophysics Workshop Pty Ltd, 3/22 Cliff St., Manly 2095, Australia\\
$^{12}$SISSA, Via Beirut 2--4, I-34014 Trieste, Italy}
\begin{document}

\date{}

\pagerange{\pageref{firstpage}--\pageref{lastpage}} \pubyear{2002}

\maketitle

\label{firstpage}

\begin{abstract}
The Australia Telescope 20\,GHz (AT20G) survey is a blind survey of the whole Southern sky at 20\,GHz with follow-up observations at 4.8, 8.6, and 20 GHz carried out with the Australia Telescope Compact Array (ATCA) from 2004 to 2008. In this paper we present an analysis of radio spectral properties in total intensity and polarisation, sizes, optical identifications, and redshifts of the sample of the 5808 extragalactic sources in the survey catalogue of confirmed sources over 6.1 sr in the Southern sky (i.e. the whole Southern sky excluding the strip at Galactic latitude $|b|<1.5^\circ$).

The sample has a flux density limit of 40 mJy. Completeness has been measured as a function of scan region and flux density. Averaging over the whole survey area the follow-up survey is 78\% complete above 50 mJy and 93\% complete above 100 mJy. 3332 sources with declination $\delta < -15^\circ$ have good quality almost simultaneous observations at 4.8, 8.6, and 20\,GHz. The spectral analysis shows that the sample is dominated by flat-spectrum sources, with 69\% having spectral index $\alpha_{8.6}^{20}> -0.5$ ($S \propto \nu^\alpha$). The fraction of flat-spectrum sources decreases from 81\% for $S_{\rm 20\ GHz}>500$\ mJy, to 60\% for $S_{\rm 20\ GHz}<100$\ mJy. There is also a clear spectral steepening at higher frequencies with the median $\alpha$ decreasing from $-0.16$ between 4.8 and 8.6\ GHz to $-0.28$ between 8.6 and 20\ GHz.

Simultaneous observations in polarisation are available for all the sources at all the frequencies. 768 sources have a good quality detection of polarised flux density at 20 GHz; 467 of them were also detected in polarisation at 4.8 and/or at 8.6 GHz so that it has been possible to compare the spectral behaviour in total intensity and polarisation. We have found that the polarised fraction increases slightly with frequency and decreases with flux density. The spectral indices in total intensity and in polarisation are, on average, close to each other, but we also found several sources for which the spectral shape of the polarised emission is substantially different from the spectral shape in total intensity. The correlation between the spectral indices in total intensity and in polarisation is weaker for flat-spectrum sources.

Cross matches and comparisons have been made with other catalogues at lower radio frequencies, and in the optical, X-ray and $\gamma$-ray bands. Redshift estimates are available for 825 sources.

\end{abstract}


\section{Introduction}

The properties of radio source populations at high ($\gtrsim10\ $GHz) radio frequency are poorly known due to the fact that large-area high-frequency surveys are very time-consuming with high sensitivity ground-based diffraction limited telescopes. The characterization of this population of radio sources is of great interest since it is dominated by the young flat-spectrum AGNs which are expected to have a strong influence on the evolution of the early Universe. Furthermore, it is crucial for the interpretation of data collected by experiments imaging the Cosmic Microwave Background (CMB). Fluctuations due to radio sources are the main contaminants of the CMB signal on scales smaller than 30 arcmin (De Zotti et al. 1999, Toffolatti et al. 2005); they need therefore to be carefully subtracted to avoid biasing the estimates of cosmological parameters. Because of the complexity and variety of radio spectra, this subtraction cannot be properly done relying on extrapolations from low frequency surveys (Massardi et al. 2009).

This has motivated large observational efforts in the last several years to conduct surveys at high radio frequency (e.g., Waldram et al. 2003, 2009; Muchovej et al. 2009; see De Zotti et al. 2010 for a review). The Australia Telescope Compact Array (ATCA)\footnote{http://www.narrabri.atnf.csiro.au/} 20 GHz (AT20G) survey, carried out from 2004 to 2008, covered all the Southern sky (6.1 sr) to a flux density limit of $\sim 40$ mJy (Murphy et al. 2010).

A Pilot Survey (Ricci et al. 2004a; Sadler et al. 2006) at 18.5 GHz was carried out in 2002 and 2003. It detected 173 sources in the declination range $-60^\circ$ to $-70^\circ$ down to 100~mJy and was used to optimize the observational techniques for the full AT20G survey.

The analysis of the 320 sources of the AT20G Bright Source Sample ($S_{20\rm GHz}>0.50\,$Jy) at declination below $-15^\circ$ was presented by Massardi et al. (2008). In this paper we extend the analysis to the full sample of confirmed extragalactic sources from the Murphy et al. (2010) catalogue.The outline of the paper is the following. \S~\ref{sec:sample} contains the description of the sample. In \S~\ref{sec:completeness} we analyze the completeness and derive source counts for the whole sample as well as for the steep- and flat-spectrum populations separately. The spectral properties in total intensity and polarisation are discussed in \S~\ref{sec:spectra} and \ref{sec:polarisation} respectively. Optical identifications, redshifts, and cross correlations with surveys in other wavebands are briefly presented and discussed in \S~\ref{sec:ID_z}. More detailed analyses are deferred to subsequent papers. The results are summarized in \S~\ref{sec:Conclusions}.

\section{Sample definition and properties} \label{sec:sample}
\subsection{Observations and sample selection} \label{sec:observations}

The AT20G survey is a blind survey using a wide band (8 GHz) analogue correlator connected (operating in the frequency range 16-24 GHz) to three of the ATCA 22m antennas.
The telescope scanned at $15^\circ$ per minute in declination along the meridian to cover $10^\circ$ to $15^\circ$ wide declination strips, with 10 mJy rms noise.
The whole Southern sky was completely covered in four observing epochs between 2004 and 2007.
A comprehensive description of the instrumental setup, the scan strategy, the map-making and source extraction techniques will be presented in a forthcoming paper (Hancock et al., in preparation).

As described in Murphy et al. (2010), candidate sources from the blind survey scans were re-observed in a compact hybrid configuration with the ATCA digital correlator with two 128~MHz bands centered at 18752~MHz and 21056~MHz and two orthogonal linear polarisations for each band, during several observing epochs between 2004 and 2008. These follow-up observations were used to confirm real detections and measure accurate flux densities and positions. The combination of the two close bands are considered as a single 256~MHz wide band centered at 19904~MHz, which is the reference frequency for our `20~GHz' observations. The main follow-up was performed soon after the blind scans, and few further runs were done to recover bad weather observations or improve the sample follow-up completeness. Almost simultaneously (i.e. within a couple of weeks) from any 20 GHz follow-up, detected sources were re-observed with an East-West configuration with two 128~MHz bands centered at 4.8 and 8.64~GHz (hereafter referred as `5' and `8'~GHz) for spectral analysis. A data-analysis pipeline was built to produce the final catalogues.
Above $\delta=-15^\circ$ the beams are very elongated for East-West ATCA arrays used for the 5 and 8~GHz observations so we did not observe the sources in this region at frequencies below 20 GHz. The primary beam FWHM is $2.4$, $5.5$, and $9.9$~arcmin at 20, 8, and 5~GHz, respectively and the angular resolution for the follow-up is typically 10 arcsec at 20 and 5~GHz and 5 arcsec at 8~GHz. The details of follow-up observations and the data reduction procedures, together with the catalogue of the surveyed sources were presented in Murphy et al. (2010).

The AT20G survey sample of followed-up sources consists of 5890 sources over 6.1 sr of the Southern sky at 20\ GHz (i.e. the whole Southern hemisphere with a Galactic cut for $|b|<1.5^\circ$). A map of the distribution of these sources is in Figure 8 of Murphy et al. (2010). Optical counterparts were obtained from the NASA Extragalactic Database (see \S\,\ref{sec:ID_z}) and in the SIMBAD database: they have been used to identify the Galactic objects in the sample. The exclusion of the $|b|<1.5^\circ$ region in fact does not guarantee that all sources in the Murphy et al. (2010) catalogue are extragalactic. The 65 sources identified as PNe were found by looking for flat spectrum sources that were slightly extended, selected according to what discussed in \S \ref{sec:high_res}. A full analysis including this data will be published separately. The 6 HII regions were identified through a search of the literature. Finally, the catalogue includes 82 sources (1.4\%) flagged as Galactic or Magellanic Cloud HII regions or planetary nebulae, which we removed from our sample, leaving 5808 sources. We kept the other sources in the Magellanic Cloud region. The sample may still contain some Galactic sources, but we believe that the residual contamination should be small because the relatively high resolution 20 GHz beam preferentially selects compact sources ($<30$ arcsec) and Galactic sources with $|b|>1.5^\circ)$ are rarely that small.

The final sample of sources used for the following analysis is composed of 5808 extragalactic sources.

\subsection{Flux density measurements}\label{sec:flux}

Flux densities of {\it point-like} sources were calculated using the triple product technique according to which the flux density of a point source is the geometric average of the visibility amplitudes in a baseline closure triangle. The flux density uncertainty is a combination of errors due to calibration and to the noise level (for further details see Murphy et al. 2010). This technique is very robust to the effects of phase de-correlation. We measured both the triple correlation and the peak amplitude in the image. These agreed very well in good weather conditions. In periods of poor phase stability the scatter between multiple observations using image amplitudes increased while the scatter in the triple correlations was unaffected. Hence, the triple product technique was used to get accurate flux density measurements for almost all the compact sources. Only 327 sources have the `poor quality' flux density flag set, due to low quality data.

The triple product technique, however, is not well-suited for extended or multiple-component sources. We have identified the extended sources by combining automatic methods and visual inspections of both images and visibility functions as described by Murphy et al. (2010). The automatic extended source identification method includes the comparison of flux density measured on short and long baselines, and the analysis of any significant deviation of the phase closure from zero which is expected for a point source. Sources with significant flux density from regions larger than 5-10 arcsec are flagged as extended using these methods. Our sample contains 308 sources in this category; thus they constitute 5\% of the total sample. For 246 of these sources the integrated flux density was obtained from the visibility amplitude measured on the shortest spacing. However, this method can fail for sources with multiple components separated by more than about 45\,arcsec, for which cases we don't have close enough short spacings to get a reliable integrated flux density. For these remaining sources the triple product is used instead as a rough estimate of the flux density of the strongest component and these sources are flagged as `poor quality' flux densities. Multiple widely separated compact components (e.g. hot spots in lobes) may be listed as distinct sources and may be misidentified (see \S\,\ref{sec:ID_z}). Sources with structure within the 2.4 arcmin 20 GHz primary beam are flagged as extended. The position in the catalogue for these sources corresponds to the peak in the image which is usually the flat-spectrum core in this high frequency catalogue, and for these sources the optical identification is correct. However, for a very small number of extended sources the strongest peak could be a hotspot in the lobes or jet and the identification might be unclear. These extended sources and refined identifications will be discussed in a future paper (Mahony et al. in preparation).

We must stress that the survey and the source detection technique is optimized for point sources, and there is bias against extended sources with angular sizes larger than about 45 arcsec. The 30m spacing used for the blind survey underestimates the amplitude by 50\% and the survey is increasingly incomplete for more extended sources.
Integrated flux densities at 20GHz have been obtained for nine very extended extragalactic sources (Burke-Spolaor et al. 2009). The flux densities determined for the seven sources included in AT20G survey were added to the final catalogue. No integrated flux densities at other frequencies are available for these sources at the same epoch; so they are not included in the spectral analysis.

Sources found to be extended at 20 GHz have been assumed to also be extended at the lower frequencies and the flux density was estimated from the shortest baselines.

\subsection{6~km visibility data}\label{sec:high_res}

Data were collected with the `6\ km' antenna which is used in conjunction with the hybrid arrays used for the ATCA during the follow-up observations. This provided 5 baselines from 4.3 to 4.5~km for 5542 (94\%) of the sources in the sample. The visibility of an interferometric measurement of a source is the Fourier transform of the brightness distribution of the source and decreases with increasing angular size, from 1.0 corresponding to a point source. The visibility is determined by the ratio of the average scalar amplitude of the 5 long ($\sim 4.5$ km) baselines to the average scalar amplitude of the 10 short (30 - 214 m) baselines. Given the typical errors in these 4.5~km visibilities, any source with measured visibility $< 0.85$ is considered to be significantly resolved. This corresponds to a size of 0.15 arcsec for a Gaussian brightness distribution or 0.23 arcsec for double peaks. We refer to these sources as `not-compact'. This information was used to separate the compact and not-compact sources discussed in the following sections.
A full discussion of these high resolution observations will be given in Chhetri et al. (in preparation).

Note that the limiting size for compact sources defined above ($\simeq 0.2$ arcsec) corresponds to a much smaller angular size than in the discussion of extended sources in the previous section which is based on the observations with the compact hybrid array alone.
The definition of `not-compact' based on the 6~km antenna visibility corresponds to sizes of more than 0.2 arcsec and contains 19\% of the 20~GHz population. As shown in \S\,\ref{sec:spectra}, this size limit divides the compact and extended source populations very cleanly and corresponds closely to the traditional classification into flat- and steep-spectrum sources.

\subsection{Polarisation}

As described in Murphy et al. (2010), the polarised flux density has been measured simultaneously with the total intensity measurements at all frequencies, and polarisation values at any frequency are given if:
\begin{itemize}
\item the polarised flux density is at least three time larger than its rms error;
\item it is larger than 6 mJy;
\item the fractional polarisation is larger than 1 per cent.
\end{itemize}
All the other cases have been considered non detections and we used as an upper limit to the polarised flux density the maximum between: a) 3 times the rms error on the polarised flux density; b) 6 mJy; and c) 1 per cent of the total flux density.

No estimation of integrated polarised flux densities was attempted for extended sources and these are not included in the polarisation analysis. This results in some bias, however to include them we would need to use the peak polarisation as given in the Murphy et al. (2010) catalogue which would result in an even worse bias because it will always be higher than the integrated polarisation which is what is measured for the unresolved sources. The only exceptions are the integrated polarisations for the sample of bright extended sources included in Burke-Spolaor et al. (2009).
\begin{figure*}
\begin{center}
\includegraphics[width=5cm]{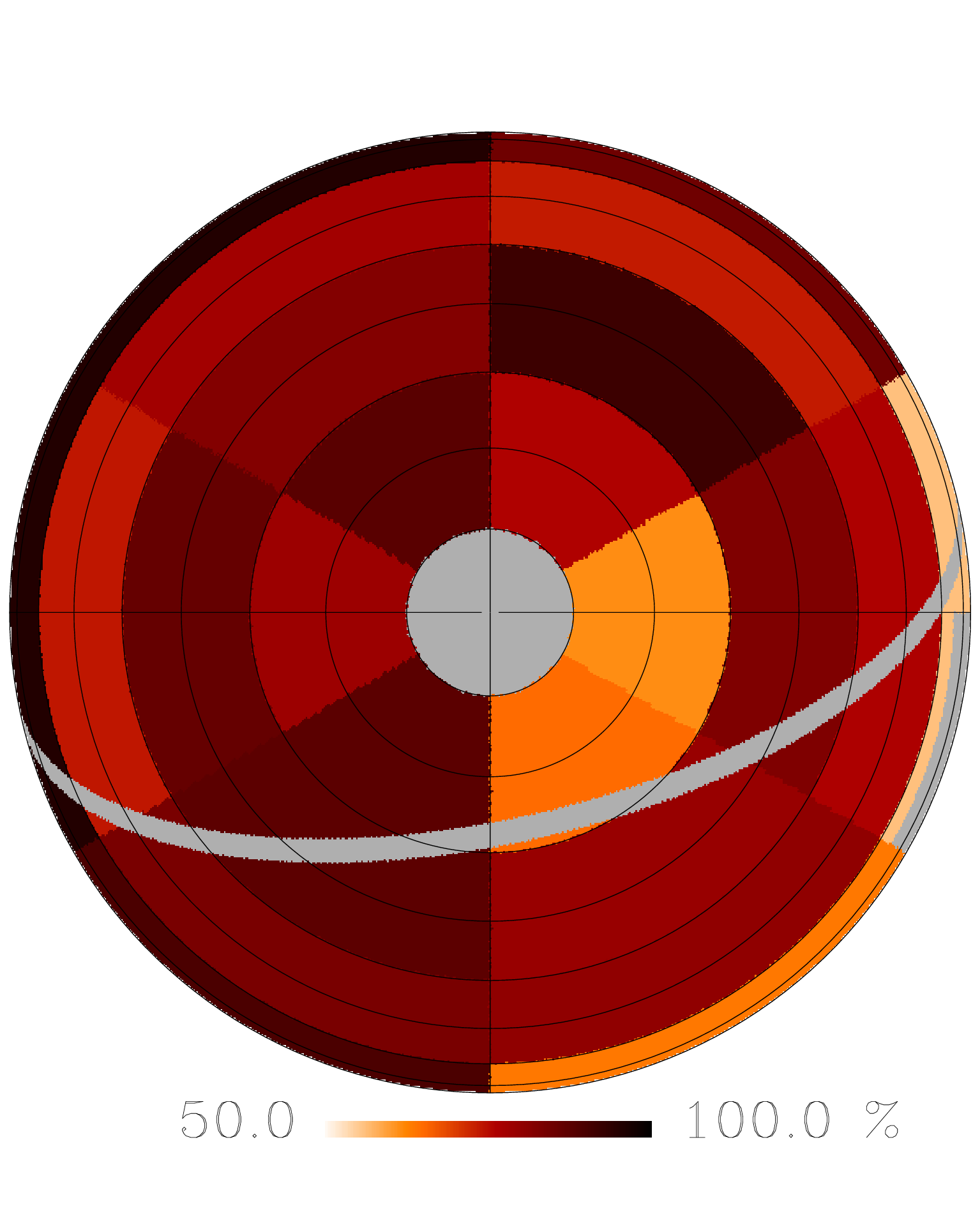}
\includegraphics[width=5cm]{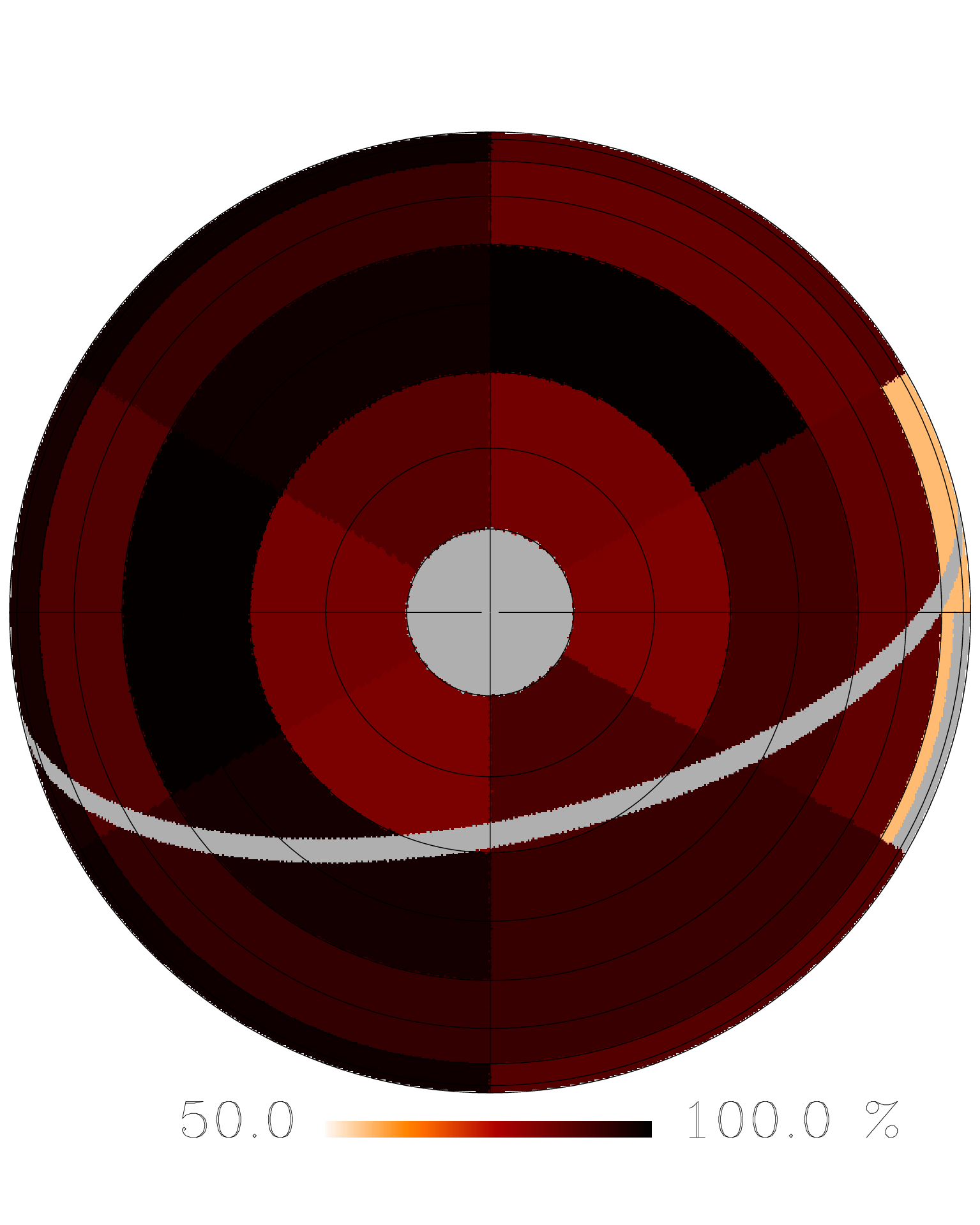}
\includegraphics[width=5cm]{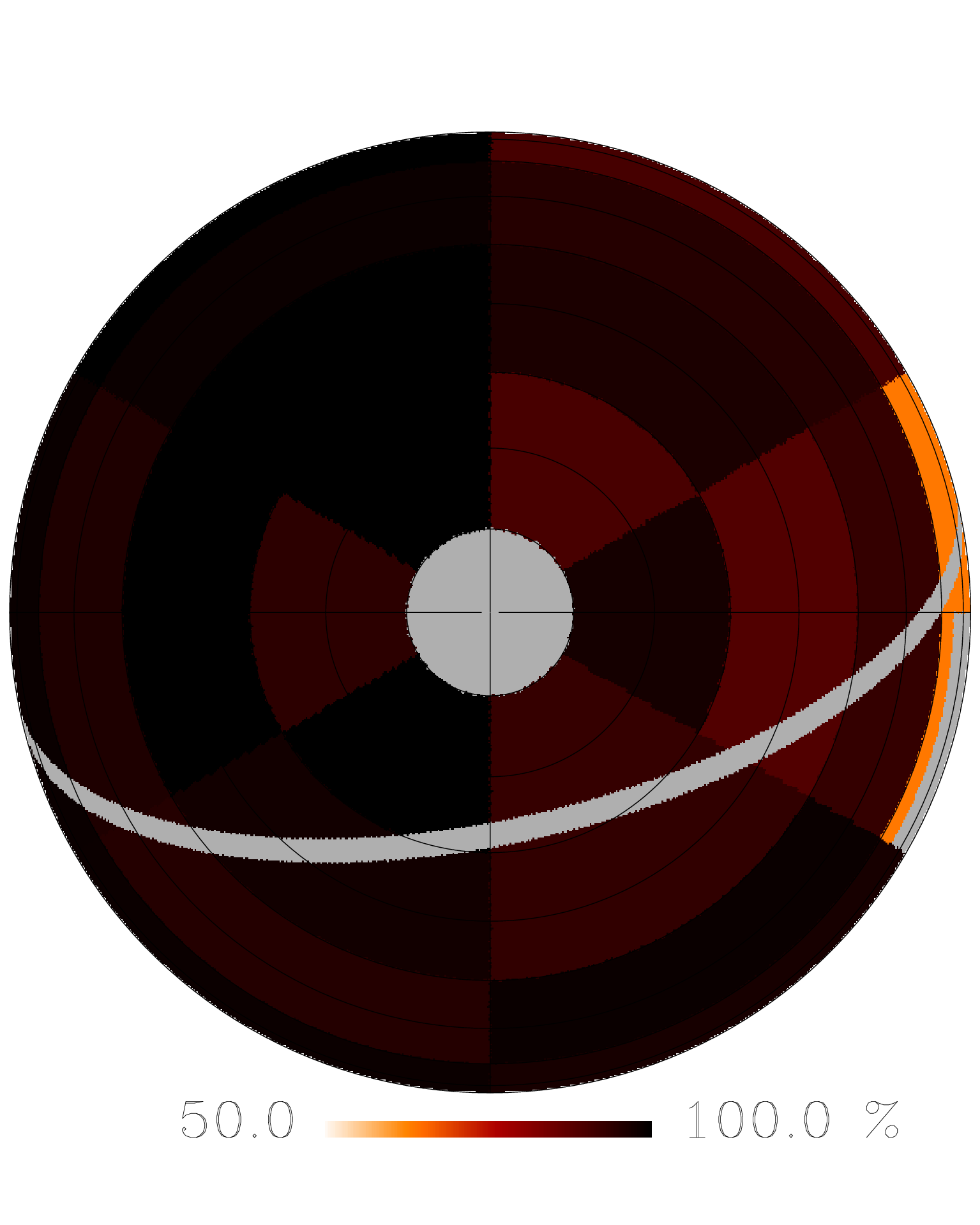}
\caption{Orthographic projection of the Southern hemisphere. Colors represent the completeness above 50, 70 and 100 mJy respectively, from left to right. RA=0 is towards the top of the figure and increase in the anticlockwise direction. The grey region corresponds to the Galactic plane region ($|b|<1.5^\circ$) and other regions not considered in the completeness analysis. The region with $\delta>-15^\circ$ and $16\hbox{h}<\hbox{RA}<18\hbox{h}$ has been discarded because the follow-up survey was seriously affected by bad weather. Also the reliability estimation techniques fail close to the South Pole and for this reason we discarded the $\delta<-80^\circ$ region in the completeness analysis and source counts.} \label{fig:comp}
\end{center}
\end{figure*}
\begin{figure}
\begin{center}
\includegraphics[width=6cm,height=8cm, angle=90]{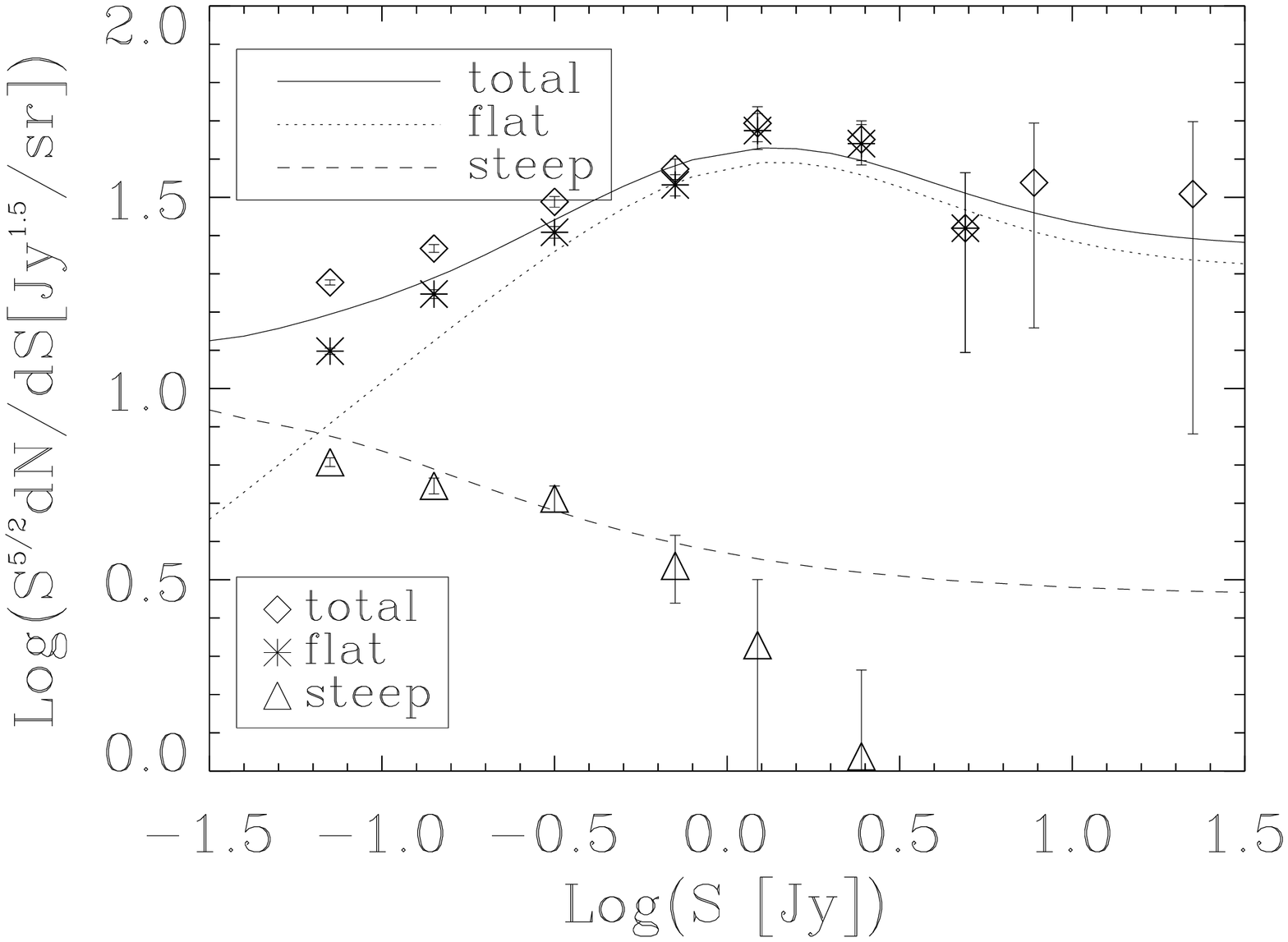}
\includegraphics[width=6cm,height=8cm, angle=90]{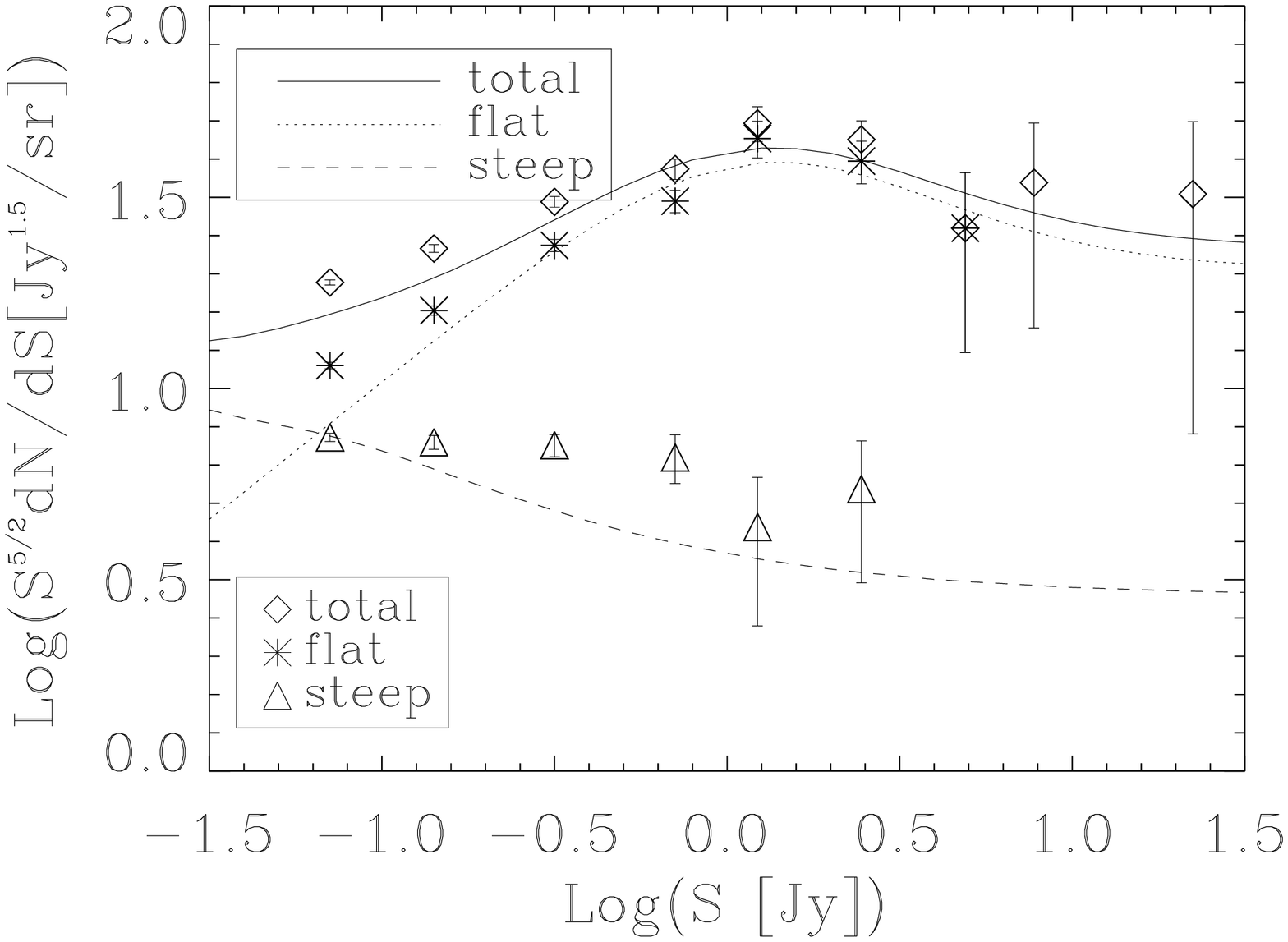}
\caption{ Euclidean normalized differential source counts for the whole AT20G sample (diamonds), for flat- (asterisks) and steep-spectrum (triangles) sub-populations. The spectral classification is based on $\alpha_{1}^{5}$ or $\alpha_{8}^{20}$ in the upper or lower panel, respectively. The counts of each sub-population were multiplied by the factor $N_{\rm tot}/(N_{\rm flat}+N_{\rm steep})$ to correct for the incomplete spectral information (see text). The lines show the predictions of the De Zotti et al. (2005) model for the whole population (solid line), and for flat- and steep-spectrum sub-populations (dotted and dashed lines respectively)}. \label{fig:srccnt}
\end{center}
\end{figure}

\section{Completeness, reliability, and source counts}\label{sec:completeness}

\subsection{Blind scan and follow-up completeness}
The follow-up observations were carried out at the position of candidate objects detected on the survey blind scans. Candidate sources were prioritized for follow-up observations in order of decreasing flux density to optimize sample completeness at high flux density levels. However, this fact constrained the flux density limit to that which could be reached in the allocated follow-up time given the variable weather conditions. As a result, different regions of the sky, observed during different follow-up campaigns have slightly different flux density completeness limits. The final sample completeness is determined by both the effectiveness of the survey source detection techniques (i.e. by the completeness of the candidate source list) and by the quality (i.e. observing strategy, weather conditions and sensitivity level) of the follow-up observations.

As discussed in Hancock et al. (2010, in preparation), and Hancock (2010) the completeness of the blind scan sample was assessed by inserting simulated sources into the data stream at the raw (uncalibrated) data level, and following these sources through all the calibration and source detection steps. With this procedure Hancock (2010) found that the blind scan sample is 97\% complete above 50 mJy and 93\% complete above 40 mJy. However, this candidate source list is only 50\% reliable in some regions of the sky at its lower flux density limits. To obtain a catalogue with essentially 100\% reliability we needed the follow-up observations but as discussed above, the follow-up observing strategy and data quality was the major limitation to the final sample completeness. To quantify the completeness as a function of flux density and region of the sky we can compare the list of confirmed sources in the follow-up catalogue with the blind scan survey source list whose completeness is known. We could not use this technique directly on the follow-up catalogue since it is affected by the weather conditions at the moment of the follow-up observations and by the efficiency of the detection algorithm that improved over time. However, we could apply this method to the blind scan survey because all the data was re-reduced with a single consistent algorithm over the entire survey. Obviously, a follow-up survey simultaneous to the blind survey could not be done after the new optimised survey reduction was completed.

Using this method the follow-up sample was found to be 78\% complete above 50 mJy, 86\% complete above 70 mJy, and 93\% complete above 100\ mJy over the whole survey area. Note that, as shown by the map in Figure~\ref{fig:comp} there are some regions with up to 95\% completeness down to 50 mJy and 100\% completeness above 70 mJy.

\subsection{Spectra}
\begin{table*}
 \caption{Euclidean normalized differential source counts ($\log S^{5/2}dN/dS [{\rm Jy}^{1/5}{\rm sr}^{-1}]$) for the AT20G sample as a whole and for the steep- ($\alpha<-0.5$) and flat-spectrum ($\alpha>-0.5$) sub-populations. Since spectral indices are different in different frequency ranges we need to specify the frequencies used for the source classification. Since spectral information is not complete, counts of each sub-population were corrected multiplying them by the factor $N_{\rm tot}/(N_{\rm flat}+N_{\rm steep})$ (see text).}
 \label{tab:srccnt}
 \begin{tabular}{cccrcrcrccrcr}
 \hline
 Flux density&&&& \multicolumn{4}{c}{$\alpha$ between 8 and 20 GHz}&\ & \multicolumn{4}{c}{$\alpha$ between 1 and 5 GHz} \\ \cline{5-8} \cline{10-13}
 range & Area & counts  & $N_{\rm tot}$ & counts& $N_{\rm flat}$ & counts& $N_{\rm steep}$ &\ & counts& $N_{\rm flat}$ & counts& $N_{\rm steep}$ \\
     (Jy)  &  (sr) & total    &                 & flat     &                & steep  &\   &                 & flat     &                & steep     \\
\hline\vspace{0.3cm}
0.05 -0.1&6.11&$1.28_{-0.01}^{+0.01}$&2349&$1.06_{-0.01}^{+0.01}$&1469&$0.87_{-0.01}^{+0.01}$&519&\ &$1.10_{-0.01}^{+0.01}$&881&$0.81_{-0.01}^{+0.01}$&454\\ \vspace{0.3cm}
0.1 - 0.2&6.11&$1.37_{-0.01}^{+0.01}$&1641&$1.20_{-0.01}^{+0.01}$&677&$0.86_{-0.02}^{+0.02}$&308&\ &$1.25_{-0.01}^{+0.01}$&739&$0.75_{-0.02}^{+0.02}$&235\\ \vspace{0.3cm}
0.2 - 0.5&6.11&$1.49_{-0.01}^{+0.01}$&909&$1.37_{-0.02}^{+0.02}$&428&$0.85_{-0.03}^{+0.03}$&128&\ &$1.41_{-0.02}^{+0.02}$&456&$0.71_{-0.04}^{+0.03}$&91\\ \vspace{0.3cm}
0.5 - 1.0&6.11&$1.57_{-0.03}^{+0.03}$&258&$1.49_{-0.03}^{+0.03}$&130&$0.82_{-0.07}^{+0.06}$&27&\     &$1.53_{-0.03}^{+0.03}$&139&$0.54_{-0.10}^{+0.08}$&14\\ \vspace{0.3cm}
1.0 - 1.5&6.11&$1.69_{-0.05}^{+0.04}$&87&$1.65_{-0.05}^{+0.05}$&52&$0.64_{-0.26}^{+0.13}$&5&\         &$1.67_{-0.05}^{+0.04}$&54&$0.33_{-0.48}^{+0.17}$&2\\ \vspace{0.3cm}
1.5 - 4.0&6.11&$1.65_{-0.06}^{+0.05}$&70&$1.59_{-0.06}^{+0.05}$&36&$0.74_{-0.25}^{+0.13}$&5&\         &$1.64_{-0.06}^{+0.05}$&40&$0.04_{-1.14}^{+0.23}$&1
\\
\vspace{0.3cm}
4.0 - 6.0&6.11&$1.42_{-0.32}^{+0.15}$&6&$1.42_{-0.32}^{+0.15}$&3&  - &                  &\       &$1.42 _{-0.32}^{+0.15}$&3&  -             & \\
\vspace{0.3cm}
6.0 - 10.0&6.11&$1.54_{-0.38}^{+0.16}$&5&              -          &      &  - &      &\ &          - &      &    - &      \\
\vspace{0.3cm}
10.0 - 50.0&6.11&$1.51_{-0.63}^{+0.19}$&3&             -          &      &  - &       &\ &          - &     &   -&      \\
\hline
\end{tabular}
\end{table*}
\begin{figure}
\begin{center}
\includegraphics[width=7cm,height=9cm, angle=90]{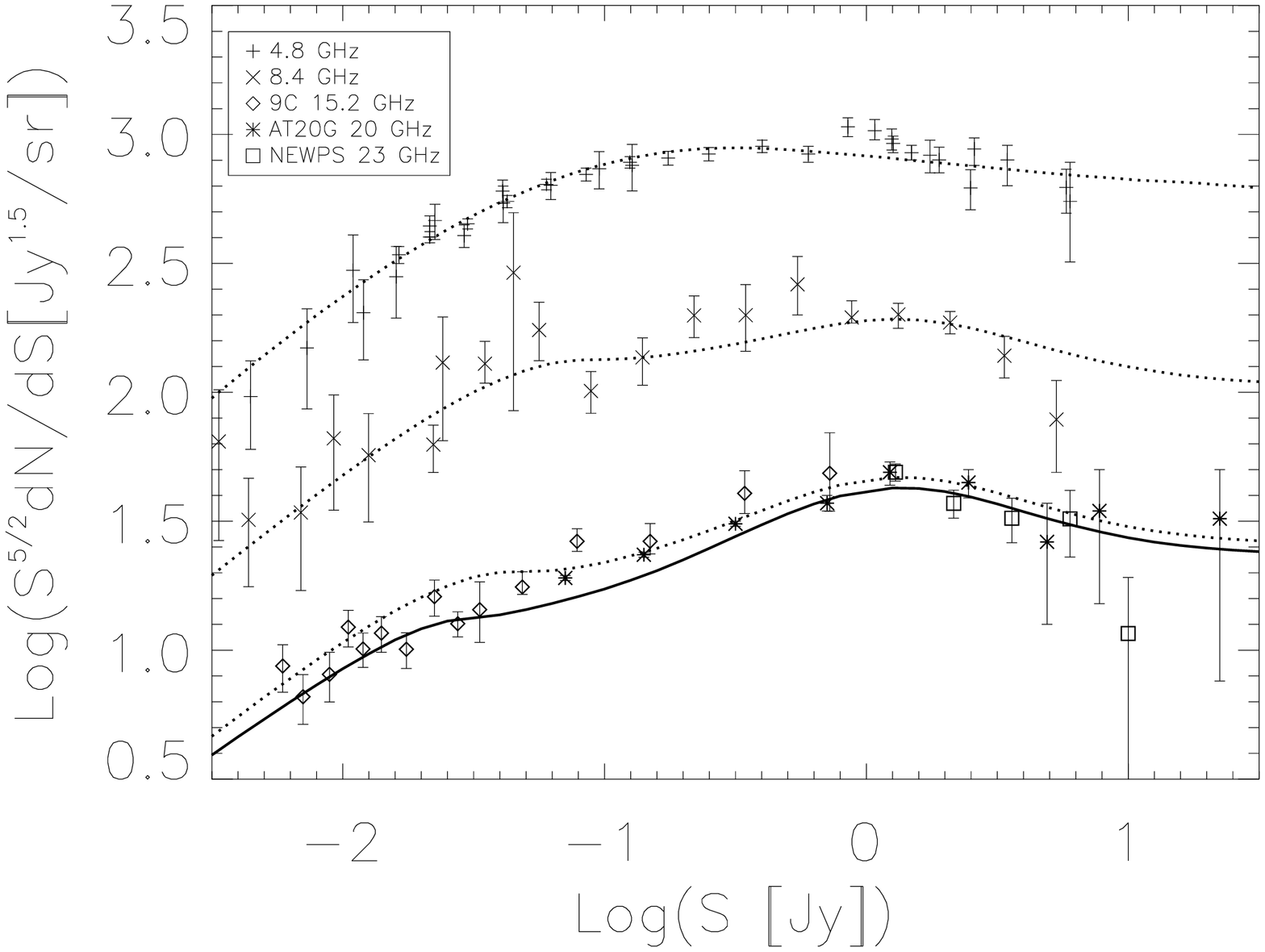}
\caption{Euclidean normalized differential source counts for the 20 GHz AT20G sample (asterisks), for the WMAP 23 GHz NEWPS sample (squares, Massardi et al. 2009), for the 15.2 GHz 9C sample (diamonds, Waldram et al. 2003, 2009), and for 8.4(x) and 4.8 (+) GHz surveys (the two latter have been multiplied respectively by a factor 10 and 100 to avoid overcrowding; see De Zotti et al. 2010 for references). Note that none of the counts have been re-scaled to match the AT20G survey frequency. The solid line shows the predictions of the De Zotti et al. (2005) model at 20 GHz, while the dotted lines are, from bottom to top, the predictions by De Zotti et al. (2005) at 15.2 and 8.4, and by Massardi et al. (2010) at 4.8 GHz.}  \label{fig:srccnt_comparison}
\end{center}
\end{figure}

As already noted in \S\,\ref{sec:observations}, no 4.8 and 8.6 GHz follow-up observations have been made above $\delta=-15^\circ$. The 3538 sources at $\delta< -15^\circ$ followed up at 20 GHz were also observed at 4.8 and 8.6 GHz within a couple of weeks and 3332 have good quality flux density measurements. The almost simultaneous observations ensure that the observed spectral behaviour is not significantly affected by variability. The AT20G pilot survey observations showed that, on a $\sim 1\,$yr timescale, the median variability index of sources selected at $\simeq 20\,$GHz is 6.9\%, and only $\sim 5\%$ of sources vary their total flux density by more than 30\% (Sadler et al. 2006).

All but 27 AT20G sources have a counterpart in the NVSS (Condon et al. 1998), SUMSS (Mauch et al. 2007), or MGPS-2 (Murphy et al. 2007) catalogues at $\sim 1\,$GHz. The 27 sources without a counterpart are discussed in Murphy et al. (2010) and in \S\,\ref{sec:spectra}. Combining our measurements with the $\sim 1\,$GHz data we computed spectral indices, $\alpha_{\nu_1}^{\nu_2}$ ($S\propto \nu^{\alpha}$), between different pairs of frequencies (see \S\,\ref{sec:spectra}), and used them to subdivide the sample into flat- and steep-spectrum sources adopting the usual boundary value $\alpha_{\nu_1}^{\nu_2}=-0.5$ (Wall 1977). Although the measurements at $\sim 1\,$GHz are far from simultaneous with those at higher frequencies, variability is not such a big problem at 1 GHz because the flux densities will be dominated by the stable extended steep-spectrum component of the source.

The main effect of variability is the bias in favour of sources in a bright phase at the selection frequency of 20 GHz. This will make some of the variable (and mostly flat-spectrum ) sources even more inverted but it would not move many sources from the steep-to the flat-spectrum category.

The error in the spectral index between pairs of frequencies depends on the flux density error and the frequency range. For the weakest sources typical rms errors are $\delta\alpha_{5}^{8}=0.14$, $\delta\alpha_{8}^{20}=0.09$, $\delta\alpha_{5}^{20}=0.06$. For $\alpha_{1}^{5}$ the formal noise errors are smaller but the actual errors may be dominated by variability and are hard to quantify. In the following analysis we use all these ranges since the observed spectral curvature is significant.

\subsection{Source counts}

Figure~\ref{fig:srccnt} shows the Euclidean normalized total counts, and also the counts of flat- and steep-spectrum sub-populations. The latter counts are corrected for the incomplete spectral information. Since the incompleteness is due to observing constraints, unrelated to source properties, we correct it by simply multiplying the counts of each sub-population by $N_{\rm tot}/(N_{\rm flat}+N_{\rm steep})$. The spectral classification is based on the 1 to 5 GHz spectral index in the upper panel, and on the 8 to 20 GHz spectral index in the lower panel. The counts of steep-spectrum sources above $\simeq 3\,$Jy are undefined because of the lack of spectral information for extended sources (see \S\,\ref{sec:flux}).
There is a clear increase of the steep-spectrum source counts based on $\alpha_{8}^{20}$, compared to those based on $\alpha_{1}^{5}$, as a result of spectral curvature steepening of high-frequency spectra.
The lines show the predictions of the De Zotti et al. (2005) model for flat-spectrum (dotted) and steep-spectrum (dashed) sources, and for the total counts. The consistency with the observed total counts is remarkably good down to $\simeq 0.3\,$Jy. At fainter flux densities the model shows a faster than observed convergence of the counts of flat-spectrum sources, and, as a consequence, under-predicts the total counts. The model counts of steep-spectrum sources are close to the observed ones (except at the brightest flux densities, where the measured flux densities are affected by resolution effects) if the spectral classification relies on $\alpha_{1}^{5}$, but are below those based on $\alpha_{8}^{20}$. This is not surprising, given that the simple power-law spectra adopted in the model are not adequate to describe the complexity of real spectra. On the other hand, Vieira et al. (2010) find that their 150 GHz selected synchrotron emitting radio sources are consistent with a flat spectral behaviour (or $\alpha \simeq -0.1$) between 5 GHz and 150 GHz, and indeed the model is found to provide a very good fit to their 150 and 220 GHz counts of radio sources. These apparently conflicting results may be reconciled by noting that a spread of spectral indices causes the effective spectral index, $\alpha_{\rm eff}$, connecting the counts at the frequency $\nu_1$ to those at the frequency $\nu_2$, to shift with frequency. In the case of a Gaussian distribution of spectral indices with dispersion $\sigma_\alpha$ and of power-law counts with differential slope $\beta$, the variation of the effective spectral index is given by (Kellermann 1964; Danese \& De Zotti 1984; Condon 1984):
\begin{equation}\label{eq:delta_alpha}
\Delta\alpha_{\rm eff} = (\beta-1)*\sigma_\alpha^2*\ln(\nu_2/\nu_1).
\end{equation}
For flat-spectrum ($\alpha > -0.5$) AT20G sources we find, for example, $\sigma_\alpha = 0.305$ between 5 and 20 GHz so that $\Delta\alpha_{\rm eff}$ is $\simeq 0.19$ for an Euclidean slope, $\beta=2.5$. For comparison, the observed mean spectral indices steepen by $\langle\alpha_8^{20}\rangle-\langle\alpha_5^8\rangle = -0.15$. This indicates that the effect of the spread of spectral indices may compensate for the steepening of the mean spectral index. Note that, strictly speaking, $\Delta\alpha_{\rm eff}$ should be compared with observations for a sample selected at 5 GHz, not at 20 GHz. The latter selection is obviously biased in favour of 20 GHz bright sources, and therefore the value of $\langle\alpha_8^{20}\rangle-\langle\alpha_5^8\rangle$ is bound to be larger (i.e. the observed steepening is bound to be smaller) than in the case of a 5 GHz selected sample. For the same values of $\sigma_\alpha$ and $\beta$, $\Delta\alpha_{\rm eff}$ between 5 and 150 GHz is $\simeq 0.47$. A more direct comparison between our results on source spectra and those by Vieira et al. (2010) is presented in \S\,\ref{sec:spec_ind}.

The De Zotti et al. (2005) model is also in good agreement with the 31 GHz counts by Muchovej et al. (2009) at flux densities around 10 mJy, i.e. fainter than those reached by the AT20G survey, although at still fainter flux density levels it under-estimates the counts. As shown by Figure~\ref{fig:srccnt}, the model predicts that, at flux density levels of a few tens of mJy, counts are dominated by steep-spectrum sources, consistent with the findings of Muchovej et al. (2009).

Figure \ref{fig:srccnt_comparison} compares Euclidean normalized differential source counts for the AT20G sample, for the WMAP 23 GHz NEWPS sample (Massardi et al. 2009), for the 15.2 GHz 9C sample (Waldram et al. 2003,2009) with lower (8.4 GHz and 4.8 GHz) frequency surveys (see De Zotti et al. 2010 for references). The predictions of the De Zotti et al. (2005) and Massardi et al. (2010) models have been overlaid for comparison. At higher frequencies the peak at high flux density due to the population of flat-spectrum AGNs is very evident and is confirmed by these observations.
\begin{figure}
\begin{center}
\includegraphics[width=7cm,height=8cm, angle=90]{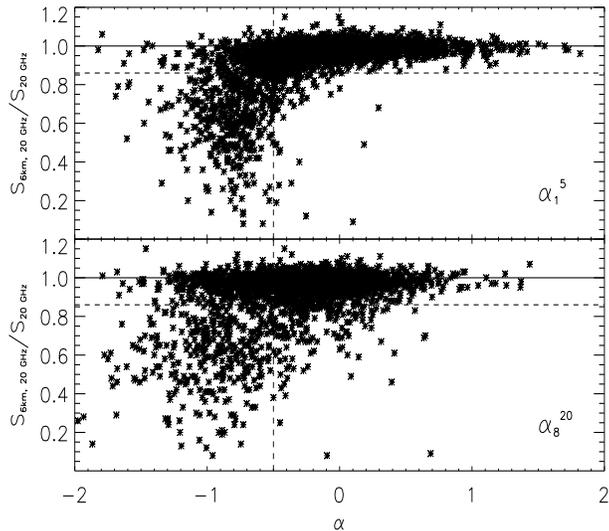}
\vspace{0.3cm}
\caption{Compactness parameter versus $\alpha_{1}^{5}$ (upper panel) or $\alpha_{8}^{20}$ (lower panel). The compactness is quantified by the ratio of the 20 GHz flux density measured with the longest baselines ($\sim 4.5$ km) to that obtained on the short baselines. Sources above a flux density ratio of 0.85 (i.e. above the horizontal dashed line) are unresolved even with the longest baseline and therefore have sizes smaller than about 0.2 arcsec. The vertical dashed line indicates $\alpha=-0.5$, used as threshold to distinguish flat- and steep-spectrum sources.} \label{fig:alpha_6km}
\end{center}
\end{figure}

\subsection{Compactness vs spectral properties}
Figure~\ref{fig:alpha_6km} shows the compactness estimated by the 6 km antenna visibility (see \S\,\ref{sec:high_res}), versus $\alpha_{1}^{5}$ (upper panel) or $\alpha_{8}^{20}$ (lower panel). A Gaussian source of 0.3 arcsec in diameter has a visibility of 0.5 on a 4.5~km baseline at 20 GHz. The unresolved sources on this plot (visibility $> 0.85$) are mostly much smaller than 0.1 arcsec so the range in distances in the sample has not smeared out this indicator of intrinsic size. 2793 of these sources also have good quality flux density measurements at the lower frequencies. There is a remarkably clean separation between steep-spectrum ($\alpha_{1}^{5}<-0.5$) extended sources and flat-spectrum compact sources. Only 3\% of the extended sources are flat-spectrum. Some of the them could be unrecognized Galactic sources with thermal free-free emission, but most are composite sources with the spectrum dominated by a flat-spectrum core but with enough emission in a jet or lobe to look extended. This class which includes upturning spectrum sources will be discussed further in \S\,\ref{sec:spectra}. From Figure~\ref{fig:alpha_6km} the time-honoured choice of $\alpha_{1}^{5}=-0.5$ as the boundary between steep-spectrum/extended and flat-spectrum/compact radio sources, originally stemming from tentative indications of a double-peaked redshift distribution in the first survey at frequency $> 1.4$ GHz (Wall 1977), asserts its solid physical basis with the spectral index acting as a proxy for source size.

The lower panel of Figure~\ref{fig:alpha_6km} shows the source compactness versus the 8-20 GHz spectral index. The compact sources show very similar behaviour but with a shift of the average spectral index to slightly steeper values and a tail extending to a steeper spectral index. The extended sources now show a much broader range in spectral index which is a combination of high frequency spectral steepening of the extended sources and an increased number of hybrid sources with upturning spectrum due to extended sources with a flat-spectrum core which dominates at high frequency.

It is also interesting that in both plots there is an apparent continuity of the compact sources across the -0.5 spectral index boundary. This strongly suggests that there is a population of steeper spectrum compact components (which includes the CSS sources) that are the tail of the compact flat-spectrum population. This is even more obvious in the lower panel, suggesting that a number of compact sources have spectral steepening at 20\ GHz as is also seen in the 95\ GHz observations (Sadler et al 2008). 12\% of the 20\ GHz survey sample are compact and steep, according to this classification, so while it's a significant extension of the population it has only a small impact on the models which divide the populations into flat- and steep-spectrum with the cut at -0.5.

\begin{figure}
\begin{center}
\includegraphics[width=8cm, angle=90]{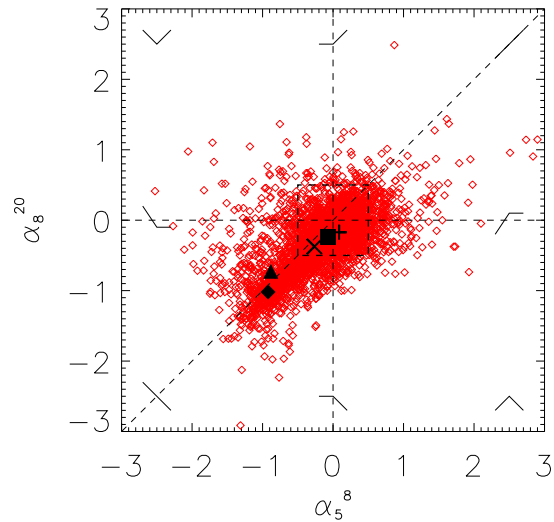}
\caption{Radio colour-colour plot for the 3332 extragalactic sources with near simultaneous observations. Sources with single power-law spectrum lie on the dashed diagonal line. The central area of the diagram collects the flat-spectrum sources. The median values have been overplotted for the extended sources (filled black diamond, see \S\,\ref{sec:flux} for details), the not-compact (filled black triangle) and compact (filled black square) sources classified according to the 6 km visibilities.
The median values of spectral indices for bright ($S_{20\rm GHz}> 500$ mJy, +) and faint ($S_{20\rm GHz}< 100$ mJy, X) are also plotted.} \label{fig:cc}
\end{center}
\end{figure}
\begin{table*}
\caption{Distribution of spectral behaviour for the 3332 sources with almost simultaneous 5, 8 and 20~GHz data and for the sub-samples with $S_{20\rm GHz}<100$ mJy, $100\leq S_{20\rm GHz}<500$ mJy, and $S_{20\rm GHz}\geq 500$ mJy. See \S\ref{sec:spectra} for details.}
  \label{tab:alphatable}
  \begin{tabular}{llcccc} \hline
                 & &    Whole sample & $<100\,$mJy& 100--500\,mJy & $\geq500\,$mJy\\
{Spectrum}& & {No. {(\%)}} & {No. {(\%)}} & {No. {(\%)}} &{No. {(\%)}}\\
 \hline
 Any    & & 3332 {(100\%) }&1544 {(46.4\%) }&1534 {(46.0\%) }&254 {(7.6\%) }\\
&&&&&\\
$\alpha_5^8>0$, $\alpha_8^{20}>0$  & \small{Inverted (I) } & 195  {(5.8\%) } &  66  {(4.3\%) } & 100 {(6.5\%) } &  29  {(11.4\%) }\\
$\alpha_5^8>0$, $\alpha_8^{20}<0$  & \small{Peaked (P) }   & 183  {(5.5\%) } & 92  {(5.9\%) } &  70 {(4.6\%) } &  21  {(8.3\%) } \\
$\alpha_5^8<0$, $\alpha_8^{20}>0$  & \small{Upturning (U) }& 102  {(3.1\%) } &  73  {(4.7\%) } &  28 {(1.8\%) } &   1  {(0.4\%) } \\
$\alpha_5^8<0$, $\alpha_8^{20}<0$  & \small{Steep (S)  }   &1086  {(32.6\%) }& 619  {(40.1\%) }& 437 {(28.5\%) }&  30  {(11.8\%) }\\
$-0.5<\alpha_5^8<0.5$ \            &                      &                  &                  &                  &                  \\
$-0.5<\alpha_8^{20}<0.5$           & \small{Flat (F) }     &1766  {(53.0\%) }& 694  {(45.0\%) }& 899 {(58.6\%) }& 173  {(68.1\%) }\\
&&&&&\\
$\alpha_8^{20}<-0.5$&\small{Steep}&1033 {(31.0\%) }&578 {(37.4\%) }&422 {(27.5\%) }&33 {(13.0\%) }\\
$\alpha_8^{20}>-0.5$& \small{Flat }        &2299  {(69.0\%) }&966  {(62.6\%) }&1112  {(72.5\%) }& 221 {(87.0\%) }\\
&&&&&\\
$\alpha_5^8<-0.5$   & \small{Steep }       & 888  {(26.7\%) }& 543  {(35.2\%) }& 324  {(21.1\%) }&  21 {(8.3\%) } \\
$\alpha_5^8>-0.5$    & \small{Flat }        &2444  {(73.3\%) }&1001  {(64.8\%) }&1210  {(78.9\%) }& 233 {(91.7\%) }\\
&&&&&\\
$\alpha_1^5<-0.5^{1}$   & \small{Steep }       & 876  {(26.7\%) }& 537  {(35.3\%) }& 322  {(21.3\%) }&  17 {(6.8\%) } \\
$\alpha_1^5>-0.5^{1}$    & \small{Flat }        &2408  {(73.8\%) }&983  {(64.7\%) }&1191  {(78.7\%) }& 234 {(93.2\%) }\\
\hline
\end{tabular}
\\
$^{1}$ Observations at 1 and 5 GHz are not simultaneous ~~~~~~~~~~~~~~~~~~~~~~~~~~~~~~~~~~~~~~~~~~~~~~~~~~~~~~~~~~~~~~~~
\end{table*}
\begin{figure*}
\begin{center}
\includegraphics[width=10cm, angle=90]{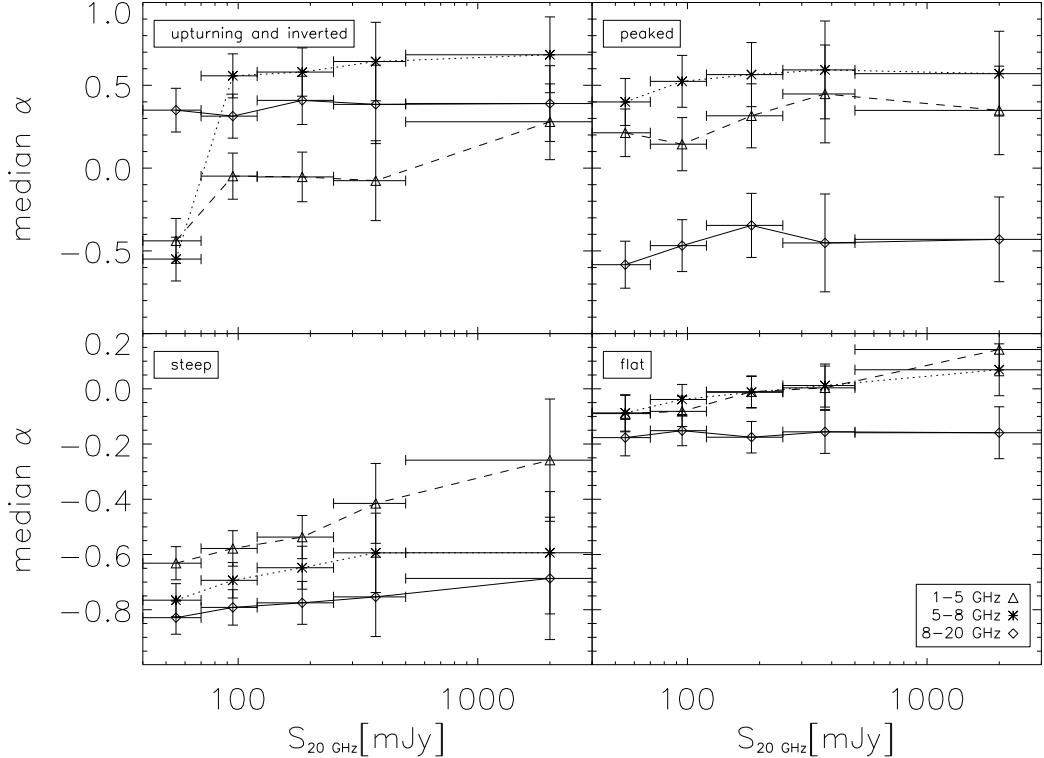}
\vspace*{0.5cm}
\caption{Median spectral indices $\alpha_8^{20}$ in several flux density bins for some spectral populations, selected combining $\alpha_8^{20}$ and $\alpha_5^8$ with the same criteria as in Table \ref{tab:alphatable} (see the text for details). The y-axis error bars have been calculated as errors on the median for a Gaussian distribution (i.e. $1.25/\sqrt{N}$).} \label{fig:alpha_s}
\end{center}
\end{figure*}

\section{Analysis of total intensity spectra} \label{sec:spectra}
\subsection{Distribution of spectral indices}\label{sec:spec_ind}

Figure~\ref{fig:cc} shows the color-color plot (Kesteven et al. 1977) comparing spectral indices at low and high frequencies. 3538 extragalactic sources, the sub-sample with near simultaneous data, were used in this analysis. 3332 sources with good flux density estimation were plotted in the Figure.

It is clear from Figure \ref{fig:cc}, and from inspection of individual spectra, that a large fraction of the AT20G sources do not have power law spectra. We investigate the effect of this spectral curvature and the way it changes with flux density in Table \ref{tab:alphatable} which summarizes the percentage of sources with different spectral behaviour in various flux density selected sub-samples. The change of population composition with varying flux density is clearly visible: the flat-spectrum population is far more dominant at the highest flux densities.

It is also clear that the ratio of the flat- to steep-spectrum sources slightly decreases with increasing frequency, indicating a mild overall steepening of the spectral index of the population.
The median spectral index over the whole sample is $\alpha_8^{20}=-0.28$.
The comparison with the median $\alpha_5^{8}=-0.16$ is a further indication of an overall steepening of the sample as the frequency increases.

We can follow this behaviour to even higher frequencies for the synchrotron emitting sources detected in the 150~GHz SPT survey. A cross-correlation of the AT20G sample with the catalog by Vieira et al. (2010) yielded 36 common sources with $S/N >5$ at both 20 and 150 GHz. Their median flux densities are 96 mJy at 20 GHz and 30.3 mJy at 150 GHz. Their median spectral index $\alpha_8^{20}=-0.12$ is somewhat flatter than that for the whole sample, as an
 effect of adding the 150 GHz selection. The mean spectral indices between 5 and 150 GHz and between 20 and 150 GHz ($\bar{\alpha}_5^{150}=-0.38$ and $\bar{\alpha}_{20}^{150}=-0.54$) are substantially steeper than the values reported by Vieira et al. (2010):  $\bar{\alpha}_5^{150}=-0.13$ for the 57 sources for which they have $\ge 5\sigma$ detections at both frequencies, and $\bar{\alpha}_{20}^{150}=-0.31$ for the top ten 150 GHz selected sources. However the mean spectral indices flatten to values close to those found by Vieira et al. ($\bar{\alpha}_5^{150}=-0.17$ and $\bar{\alpha}_{20}^{150}=-0.30$) if we emphasize the 150 GHz selection by taking the sub-sample with $S_{150\rm GHz} \ge 50\,$mJy. We then conclude that the source spectral properties are strongly affected by the selection frequency. AT20G sources, selected at 20 GHz, mildly steepen between 5 and 20 GHz, while SPT sources selected at 150 GHz exhibit a very flat spectral index across the frequency range 5--150 GHz.

The only remaining significant difference with Vieira et al. concerns $\bar{\alpha}_{5}^{20}$; for the top ten 150 GHz selected sources they find $\bar{\alpha}_{5}^{20}=0.25$, to be compared with $\bar{\alpha}_{5}^{20}=-0.14$ for the full AT20G/SPT sample, and to $\bar{\alpha}_{5}^{20}=0.03$ for the sub-sample with $S_{150\rm GHz} \ge 50\,$mJy. This difference is likely due to the use of different 5 GHz data:  we used AT20G 5 GHz fluxes, while they used their own 5 GHz measurements, and the two sets of measurements were carried out at different epochs.

\begin{figure}
\begin{center}
\includegraphics[width=8cm, height=8cm, angle=90]{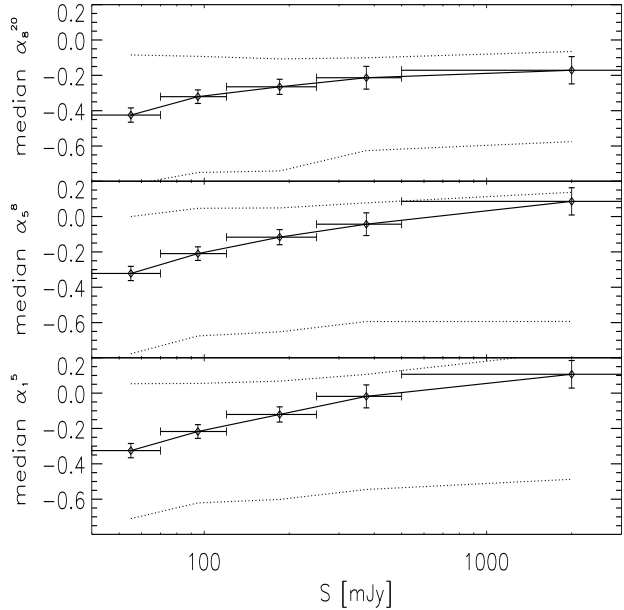}
\vspace*{0.5cm}
\caption{Median values of $\alpha_8^{20}$, $\alpha_5^{8}$, and $\alpha_1^{5}$ (diamonds and black solid line, from top to bottom panel) for different flux density bins. The y-axis error bars have been calculated as in Figure \ref{fig:alpha_s}. The dotted lines show the values of the first and third quartiles of the distributions in each flux density bin.} \label{fig:alpha_S20_distr}
\end{center}
\end{figure}

\begin{figure}
\begin{center}
\includegraphics[width=6cm, angle=90]{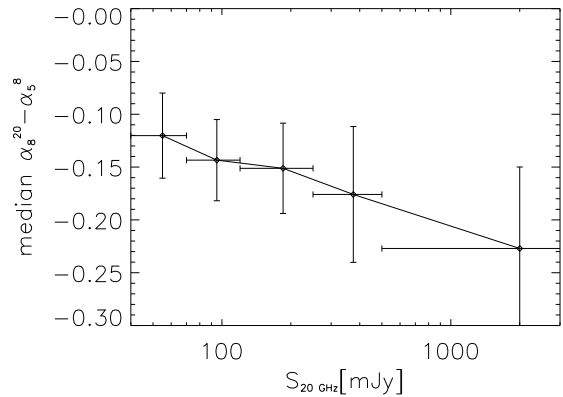}
\caption{Median values of the spectral curvature $\alpha_8^{20}-\alpha_5^8$ for different flux density bins.} \label{fig:delta_alpha_S20}
\end{center}
\end{figure}

By cross-matching the AT20G sample with NVSS, SUMSS, and MGPS-2, respectively at 1.4 and 0.834 GHz, we can extend the comparison to lower frequencies {in Table \ref{tab:alphatable}}. The median is $\alpha_1^{5}=-0.16$. However, the $\sim 1$ GHz observations are not coeval with the AT20G ones and for this reason spectral indices may be affected by variability, and the average spectral index may be biased towards high values because of the high-frequency selection. The comparison of the spectral indices over the three frequency ranges are plotted against the 20 GHz flux density in Figure \ref{fig:alpha_s} for the four spectral populations. This shows that the overall spectral steepening is more prominent for the brightest sources (see also Figure \ref{fig:alpha_S20_distr} and \ref{fig:delta_alpha_S20}) that are typically flatter than the fainter sources (at any frequency): the fraction of flat-spectrum sources is $87\pm 6\%$ if $S_{\rm 20 GHz}>500$ mJy, while it is $63\pm 2\%$ if $S_{\rm 20 GHz}<100$ mJy.
\begin{figure}
\begin{center}
\includegraphics[width=6cm, angle=90]{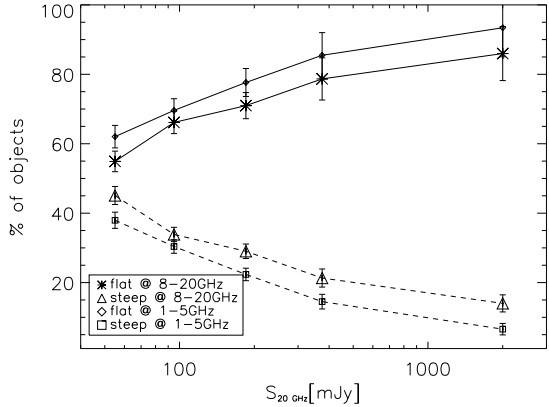}
\caption{Percentage of flat- (asterisks) and steep- (triangle) spectrum sources defined between 8 and 20 GHz together with flat- (diamonds) and steep- (squares) spectrum sources defined between 1 and 5 GHz for different flux density bins.} \label{fig:ratio_sf_S}
\end{center}
\end{figure}

\begin{table*}
\caption{UIS found in the AT20G sample with 5-8 GHz spectral index $> +2.5$.} \label{tab:UIS}
\begin{tabular}{lccccccclc}
\hline
name           &$S_{20\rm GHz}$&$S_{8GHz}$& $S_{5\rm GHz}$ & $S_{\rm NVSS}$& $\alpha_8^{20}$&$\alpha_5^8$ &err($\alpha_5^8$) &   comments\\
&  [mJy] &  [mJy] &  [mJy] &  [mJy] &&&&\\
\hline
J070949-381152 & 86$\pm$3& 33$\pm$3 &6$\pm$1  &$<2.5$&+1.1  & +3.0  &  0.3    & UIS, 2MASS galaxy with K=12.7mag \\
J111246-203932 & 78$\pm$5& 30$\pm$2 &6$\pm$1  &$<2.0$&+1.1  & +2.9  &  0.3    & UIS, No obvious optical ID \\
J143608-153609 & 113$\pm$6&53$\pm$3 &10$\pm$1 & 8.0  &+0.9  & +3.0  &  0.2    & UIS, No obvious optical ID\\
\hline
\end{tabular}
\end{table*}

\subsection{The different spectral types}\label{sec:spec_type}

In Figure~\ref{fig:alpha_s} we compare the change in spectral properties with flux density. The average difference in the median spectral index curves in this plot is a simple selection effect for the different spectral types, but the variation with flux density is not and shows interesting trends. At high frequencies the spectrum of the flat-spectrum sources is essentially independent of flux density while the spectrum of the steep-spectrum sources becomes steeper at lower flux densities at all the frequencies. This effect may be partly caused by some migration of sources between the flat- and the steep-spectrum source cut since the steep-spectrum sources are an increasing fraction of the low flux density sample. The sources with inverted and upturning spectra behave as expected if they are composite sources with a flat-spectrum core and steep-spectrum jets or lobes. At high frequency the core dominates and they have no dependence on flux density while their low frequency spectrum behaves like the steep-spectrum class of sources. Apart from the larger range of spectral differences from high to low frequencies the peaked sources have little changes of spectral indices with flux density and look more like the flat-spectrum source population.

The distribution of spectral indices is asymmetric with a tail towards steep spectra (see Figure \ref{fig:alpha_S20_distr}). The asymmetry is more clearly observable at the brightest flux densities and at the lower frequencies (where the difference of spectral behaviour between faint and bright sources is more significant). This is a clear indication that extrapolations of flux densities from a low to a high frequency using only the mean or the median value for the spectral index are incorrect. A more correct approach should consider a flux density dependent spectral index distribution. Furthermore, extrapolations performed over large frequency intervals should also consider the spectral steepening which is significant at all flux density levels but most pronounced at high flux densities (see Figure \ref{fig:delta_alpha_S20}).

The difference between the predicted and the measured ratio of flat- and steep-spectrum sources at the lower flux densities could be caused by the pair of frequencies used to make the separation. This would not change in the models which assume only power law spectra. Figure \ref{fig:ratio_sf_S} shows the fractions of flat- and steep-spectrum sources selected at the two extremes of our frequency range. There is a similar strong flux density dependence with an offset between the low and high frequency ranges.

\subsection{Ultra inverted spectrum sources}

Murphy et al. (2010) introduced the class of Ultra-Inverted Spectrum (UIS) radio sources which have a spectral index $\alpha_5^{20}> 0.7$ and usually no detectable low frequency counterpart. In Fig \ref{fig:cc} we see an even more extreme attribute of a
subsample of these sources. 4 sources in the sample are listed with a 5-8 GHz
spectral index $\alpha_5^{8}> +2.5$. Three of these with good quality flux density estimates are listed in Table \ref{tab:UIS} with their spectral index at 5-8 and 8-20 GHz and 1 sigma errors.
These sources {are very weak at the lower frequencies} so the errors are
relatively large.

All sources are just consistent with the maximum possible
synchrotron self absorption spectral index of +2.5 at the lower frequencies but are
not consistent with the maximum free-free absorption spectrum of +2.0. This steep
self absorption spectrum suggests that these sources have been confined in a
particularly dense environment, or are very young.

\section{Polarisation properties} \label{sec:polarisation}

\begin{figure}
\begin{center}
\includegraphics[width=7cm, angle=90]{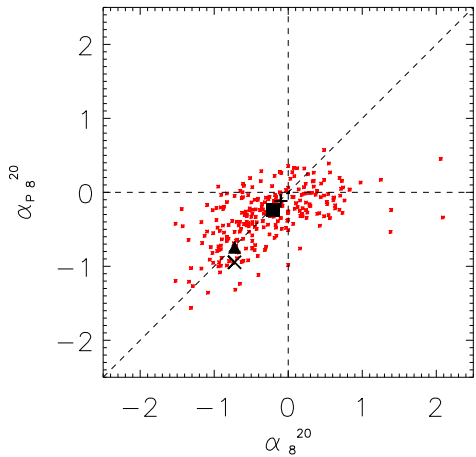}
\caption{Comparison of 8--20 GHz spectral indices in polarisation (y-axis) {and in total intensity (x-axis)}. Filled and open larger symbols are {median values as described in} Figure \ref{fig:cc}.} \label{fig:alphaPS}
\end{center}
\end{figure}

Polarisation was detected at 20 GHz for 768 sources, 467 of which also have simultaneous detections at 5 and/or 8 GHz. For each of these sources it is possible to compute the spectrum of the polarised emission. We constructed the matrix in Table \ref{tab:matrix} by comparing the total intensity and the polarisation spectral behaviours for the sources with almost simultaneous total intensity observations and polarisation detection at 5, 8 and 20~GHz. Although the spectral indices in total intensity and in polarisation are correlated (see Figure \ref{fig:alphaPS}), for many sources the spectral shape in polarisation is substantially different from that in total intensity. Flat (F) sources, which are, on average, brighter, also have a higher polarized flux density (although not a higher polarisation {\it degree}) and this translates into a larger number of detections, but the spectral shape of the detected signal is not easily predictable. The behaviour of the steeper spectrum power law sources is more predictable. Figure~\ref{fig:alphaPS} indicates that sources with power law spectra in total intensity tend to preserve the same spectral shape in polarisation, whereas flat spectra in total intensity may correspond to a large range of possible spectral shapes in polarisation.

\begin{table}
 \caption{Matrix of the number of sources classified according to both the total intensity and polarisation spectral behaviour. The rows are the spectral shape in polarisation, the columns the spectral shape in total intensity.
 The spectral types are defined in Table~\ref{tab:alphatable}.} \label{tab:matrix}
 \begin{tabular}{|l|c|c|c|c|c|}
 \hline
 {S $\rightarrow$}&(U)&(I)&(F)&(P)&(S)\\
 {Pol. $\downarrow$ } &&&&&\\
 \hline
 Upturning (U)  & 1 & 1 & 15 & 0 & 6 \\
 Inverted (I)  & 0 & 7 &
 25 & 2 & 1 \\
 Flat (F)  & 0 & 1 & 65 &
 3 & 10 \\
 Peaked (P)  & 0 & 2 & 43 & 3 & 17 \\
 Steep (S)  & 0 & 0 & 25 & 0 & 75 \\
 \hline
\end{tabular}
\end{table}

Low frequency depolarisation will move sources above the diagonal in the matrix but this seems to be a minor effect in the 5-20 GHz frequency range given the large number of sources that have steep-spectrum in both total intensity and in polarisation. Low frequency depolarisation is seen as a small excess of sources above the power law line in the steep-spectrum region of Figure \ref{fig:alphaPS}.

The half of the matrix below the diagonal in Figure \ref{fig:alphaPS} includes a class of sources that show decreased polarised flux density at the higher frequencies and includes a distinct group of sources with positive spectral index in total intensity and negative spectral index in polarisation. These sources may be composites with a steeper spectrum polarised component and a less polarised flat-spectrum core that dominates at higher frequency.

To calculate the median fractional polarisation at 20 GHz we have exploited the Survival Analysis techniques and the Kaplan-Meyer estimator (as implemented in the ASURV code, LaValley et al. 1992). The large number of upper limits (83 percent of the sample) at 20 GHz makes any Survival Analysis over the whole sample unreliable. To get meaningful results we need to restrict ourselves to the 1059 sources with $S_{20\rm GHz}>250\,$mJy, for which the number of upper limits (526) is $< 50$ percent of the sample. If only the robust detections are used the median fractional polarisation degree is 2.6 percent. For these sources the Survival Analysis yields a mean polarisation degree of 2.7 percent, in perfect agreement with the mean polarisation degree at 22 GHz measured by Jackson et al. (2010) for a complete sample of extragalactic sources stronger than 1 Jy in the 5-year WMAP catalogue ($2.7\pm 0.2$ percent) and with the median polarisation degree at 18.5 GHz of southern flat-spectrum sources in the K\"uhr et al. (1981) 1 Jy sample, found to be, again, 2.7 percent (Ricci et al. 2004b).

There is no evidence of a higher fraction of integrated polarised flux density at low flux density levels.

We can also use the compactness classification based on the long baselines to look at the polarisation characteristics of the compact and extended sources. This confirms the observation that most steep-spectrum extended sources have the same spectral behaviour in both total and polarised intensity {see Fig. \ref{fig:alphaPS}). If we consider only the two groups of compact flat-spectrum and compact steep-spectrum sources the mean of the {\it detected} values is 2.9\% and 3.8\%, respectively.

The ratio of fractional polarisation at different frequencies is a measure of depolarisation. The median of the ratio $m_{5}/m_{20}$ is 0.83 indicating on average a 17\% depolarisation at 5 GHz at any flux density level.
\begin{figure}
\begin{center}
\includegraphics[width=6cm, angle=90]{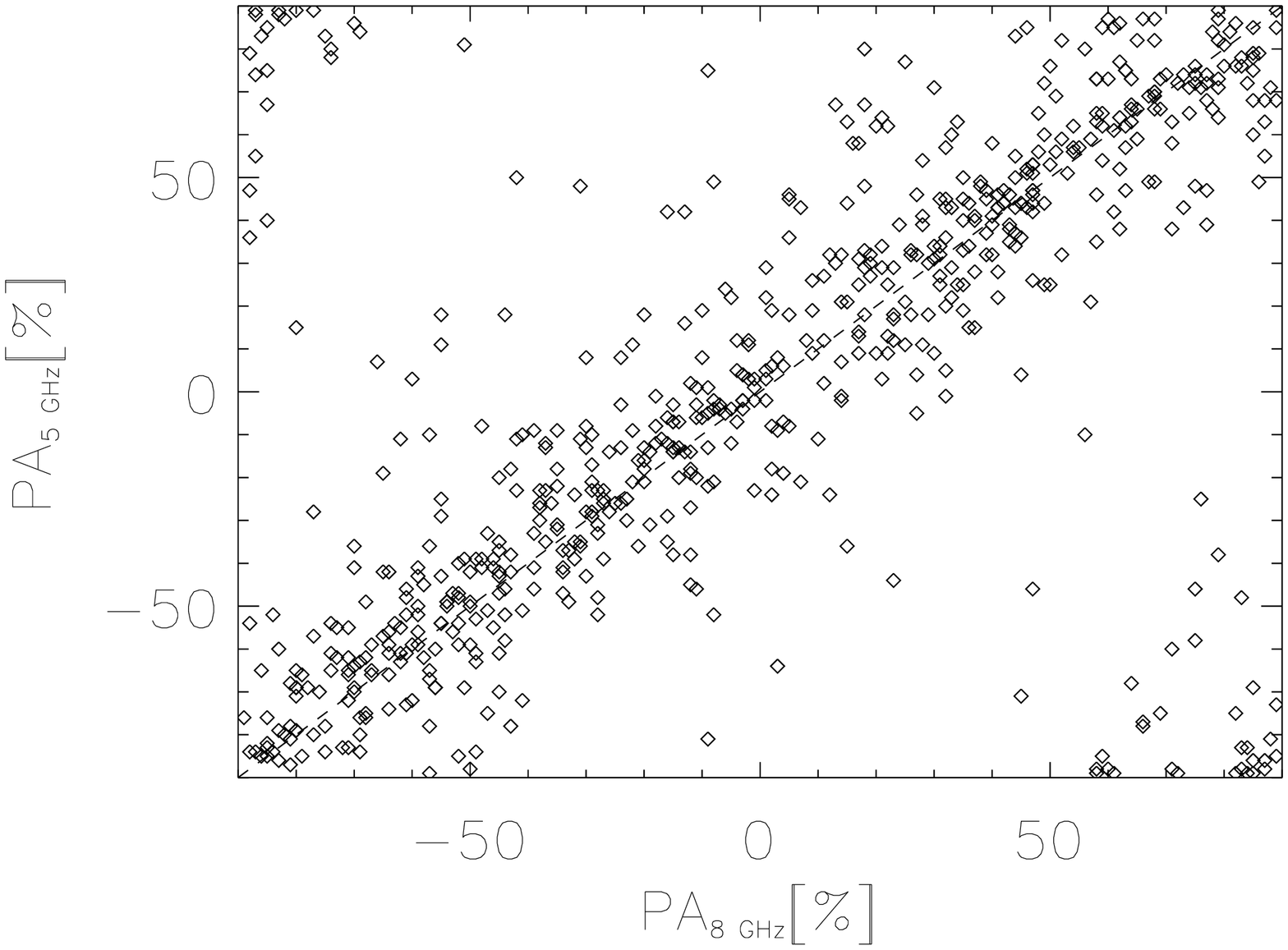}
\includegraphics[width=6cm, angle=90]{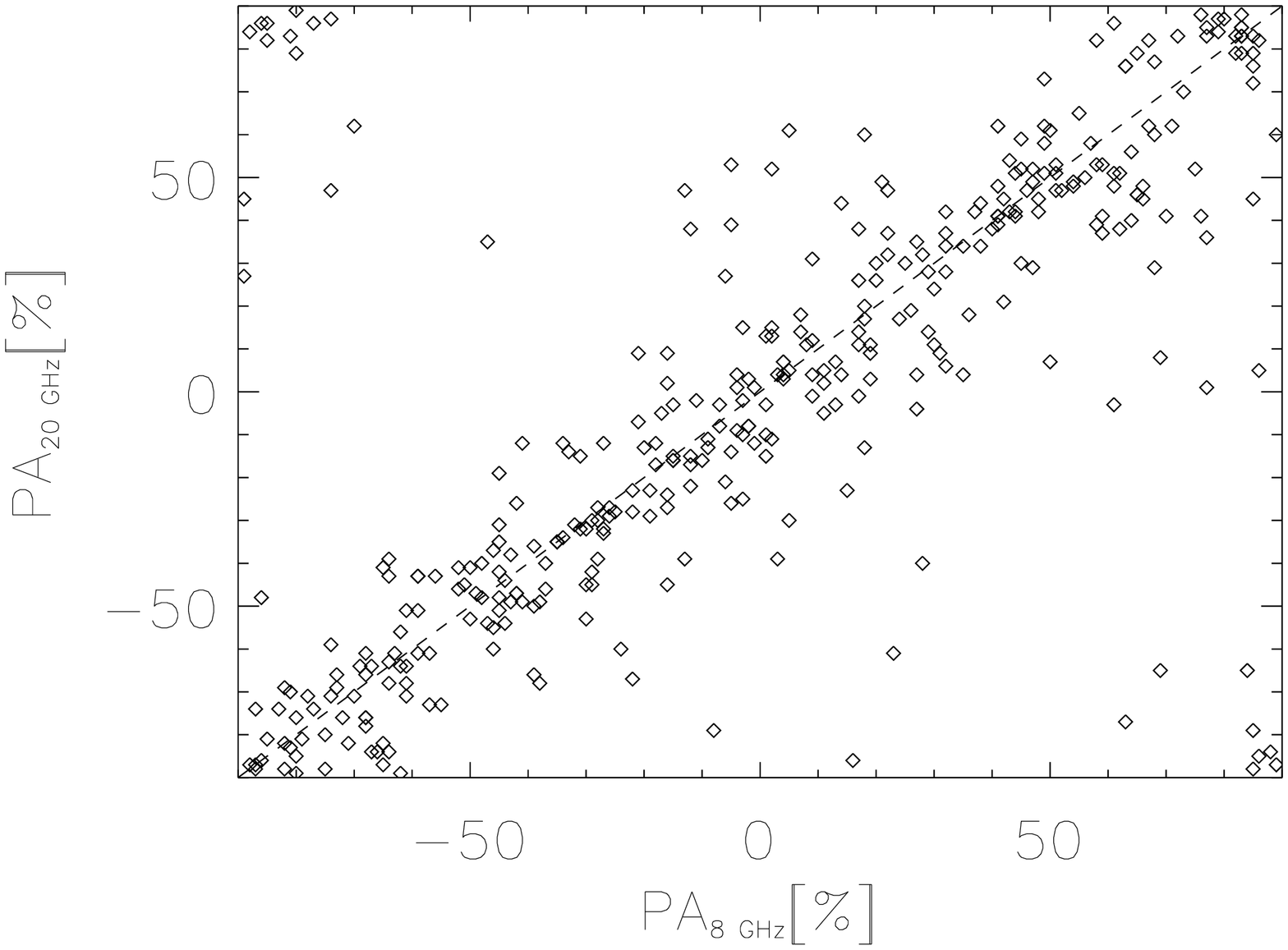}
\caption{Comparison of polarisation angles at 5 and 8 GHz (upper panel) and at 8 and 20 GHz (lower panel).} \label{fig:PA_8_20}
\end{center}
\end{figure}

The polarisation angle is quite similar at all the frequencies (see Figure \ref{fig:PA_8_20}).
However, the position angle differences have errors which are too large to determine meaningful rotation measures and the scatter is consistent with the expected intrinsic dispersion of rotation measure observed at lower frequencies.

15 sources show polarisation fraction larger than 5 percent with 20 GHz flux density larger than 1 Jy. The source with the brightest polarised flux density is the well-known blazar 3C279, with $P_{20}=$1.1 Jy and $S_{20\ GHz}=$20 Jy. The source with the largest fractional polarisation is AT20GJ195817-550923 which is 19 per cent polarised with $S_{20\ GHz}=$0.058 Jy

\section{Cross-matches in other spectral bands} \label{sec:ID_z}
\subsection{Lower radio frequencies}
Cross matches with catalogues at 1.4 GHz(NVSS) and 0.843 GHz (SUMSS-MGPS-2) have been discussed in \S\,\ref{sec:spectra} and in Murphy et al. (2010). Such comparisons demonstrated that the dominant population changes with frequency and that extrapolations assuming power laws behaviour from GHz frequencies cannot reconstruct the 20 GHz population.

Only 0.4\% of the AT20G sources have no low frequency counterpart. They have non simultaneous spectral index between 1 and 20 GHz larger than 0.7 and mostly belong to the inverted spectra sources group in the upper right quadrant of Figure \ref{fig:cc}. These belong to the class of sources with $\alpha_5^{20} > 0.7$ that in Murphy et al. (2010) have been defined as `Ultra Inverted Spectra sources'. 1.2\% of the AT20G sources belong to this class.

\subsection{Optical identifications and redshift}
Optical counterparts for the 4932 AT20G sources with $|b|>10^\circ$ were found by cross-matching the radio positions with optical positions in the SuperCOSMOS database (Hambly et al.\ 2001). Optical identifications were chosen to be the closest optical source to the radio position within a 2.5 arcsec radius and with a B magnitude brighter than 22, which is the SuperCOSMOS completeness limit (see Sadler et al. 2006). This resulted in 2958 (60$\%$) sources with optical identifications. The SuperCOSMOS classifications revealed that 1762 (60$\%$) are point-like and 1172 (40$\%$) are extended (for the remaining sources no classification is available). The simple position-based optical identification procedure is good for point sources but 2-3\% of the sources with more complex radio structure are likely misidentified. A more detailed analysis of the optical identifications including source structure will be made in a future paper (Mahony et al., in preparation).

The fraction of sources with optical identifications in the full AT20G sample is less than is seen in the AT20G Bright Source Sample (Massardi et al. 2008). This is due to the decrease in flux density limit from 500\,mJy in the BSS to 40\,mJy for the full sample and corresponds to the increase in the number of galaxies as the 20 GHz flux density decreases.

Redshift information for 825 (28$\%$ of those with optical identifications) sources was obtained by searching the NASA Extragalactic Database (NED\footnote{http://nedwww.ipac.caltech.edu/}). Of these, 443 are classified as point-like in the optical and 379 as extended. The median redshifts are 1.25 and 0.25 for each of these classes respectively. Similar fractions, 76\% and 71\%, of the point-like and extended sources respectively have flat spectra between 8 and 20 GHz.

\begin{figure}
\begin{center}
\includegraphics[width=7cm, angle=90]{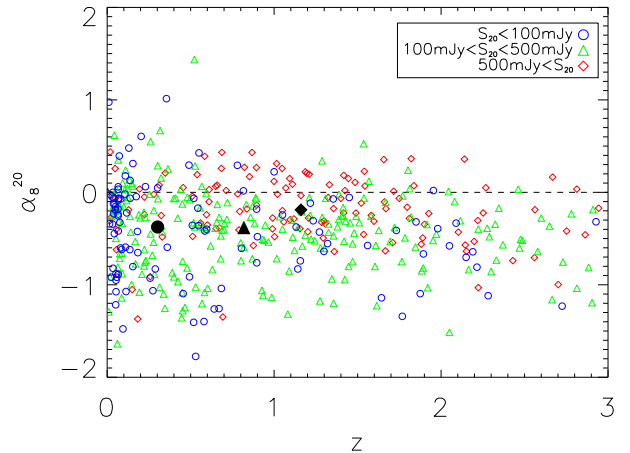}
\caption{Spectral indices $\alpha_8^{20}$ versus redshift for different flux density intervals: $S_{20\rm GHz}>0.5$ Jy, diamonds; $0.5>S_{20\rm GHz}>0.1$ Jy, triangles; $S_{20\rm GHz}<0.1$ Jy, circles. Filled symbols correspond to median values for each flux density selected sub-sample.} \label{fig:alpha_z}
\end{center}
\end{figure}

Spectral index versus redshift is plotted in Figure \ref{fig:alpha_z} for various flux density ranges. The most obvious effect is that most of the bright sources are flat-spectrum high redshift sources, while the closest sources are fainter and include a larger population of steep-spectrum sources. With redshifts for only 14\% of the AT20G sources there are a number of strong selection effects. The strong flat-spectrum AGNs are bright quasars with easily measured redshifts. The galaxies in the sample are fainter and redshifts are strongly biased to the closer galaxies. For nearby sources ($z$ $<0.2$) the redshift surveys are providing redshifts and some complete subsamples are available. Sadler et al. (in preparation) are analysing the 203 AT20G sources in the 6dF Galaxy Survey. In the following discussion we restrict our comments to redshift effects which are less influenced by sample completeness.

Figure~\ref{fig:alpha_z} indicates that the flat-spectrum population shows a steepening of the $\alpha_8^{20}$ spectral index at high $z$ but the steep-spectrum population does not show any significant change with redshift. As illustrated by Fig.~\ref{fig:dalpha_z}, however, both populations show a slight increase with redshift of the spectral curvature at lower frequencies ($\alpha_8^{20}-\alpha_5^{8}$).  This behaviour is similar for all flux density ranges and is the expected effect of the K correction on a population of sources with curved spectra. The redshift distribution is similar for flat- and steep-spectrum sources. We have redshift measurements for 298 flat-spectrum and 147 steep-spectrum sources; the median values are 0.83 and 0.93, respectively.

\begin{figure}
\begin{center}
\includegraphics[width=7cm, angle=90]{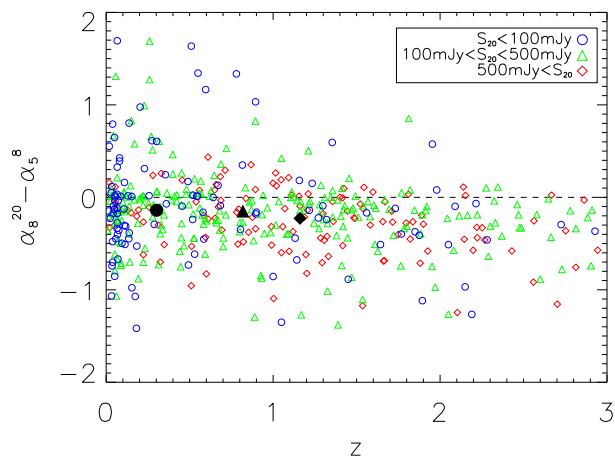}
\caption{Variation of spectral index $\alpha_8^{20}-\alpha_5^{8}$ vs redshift. Symbols are the same as in the previous image.} \label{fig:dalpha_z}
\end{center}
\end{figure}

\subsection{Cross-matches in the X-ray band}
The ROSAT All-Sky Survey Bright Source Catalogue (RASS-BSC, Voges et al. 1999) includes 8896 X-ray sources in the Southern sky with $|b|>5^\circ$. 280 sources from the whole AT20G sample (4.6\%) have a counterpart within 10 arcsec. The median 20 GHz flux density for the X-ray detected sample is 189 mJy. The median $\alpha_8^{20}$ for the 161 extragalactic sources with data on the three AT20G follow-up frequencies is -0.20; the median $\alpha_5^{8}$ of the same sources is $-0.06$. Only 76 sources with flux density larger than 189 mJy have a polarisation detection. The median fractional polarisation is 2.7\%, equal to the average calculated over the whole sample.
There is no clear relation between X-ray and radio flux densities in Figure \ref{fig:rosatS20}. However, the X-ray detection rate for AT20G survey sources increases with increasing 20 GHz flux density, as shown in Figure \ref{fig:numdet_rosatS20}. Here we binned the AT20G catalogue in flux density and plotted the percentage of those that were detected in the ROSAT dataset for each bin. There is a clear trend in the percentage of detections, implying that the brightest sources have higher probabilities of showing a counterpart in the X-ray band and, as a consequence, a relation between the mechanisms that generate the flux densities in the two bands.

\begin{figure}
\begin{center}
\includegraphics[width=6cm, angle=90]{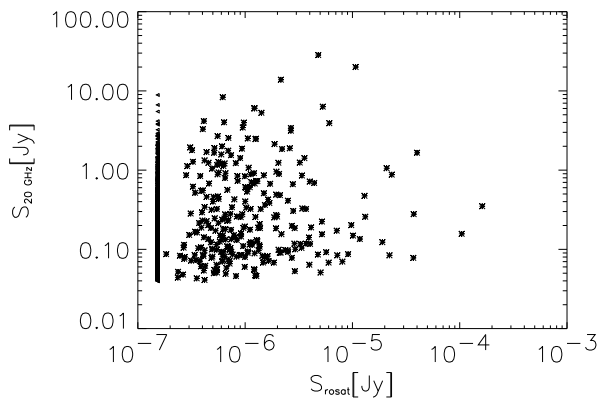}
\caption{Comparison of 20 GHz flux densities with X-ray flux densities from the ROSAT-BSC (Voges et al. 1999).} \label{fig:rosatS20}
\end{center}
\end{figure}

\begin{figure}
\begin{center}
\includegraphics[width=6cm, angle=90]{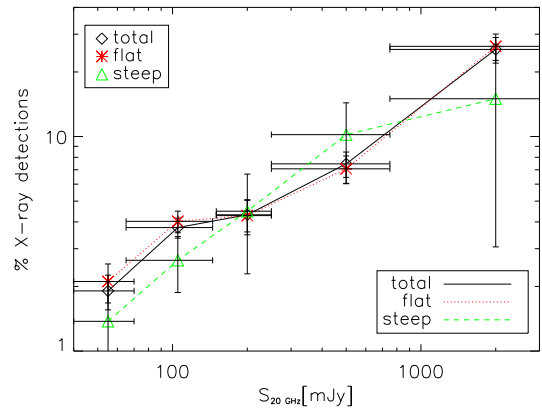}
\caption{Percentage of X-ray detections at varying 20 GHz flux densities for the whole sample (diamonds), and for flat- (asterisks) and steep-spectrum (triangles) sources.} \label{fig:numdet_rosatS20}
\end{center}
\end{figure}

\subsection{Cross-matches with the $\gamma$-ray band}
These results are presented in more detail in Mahony et al. (2010) but we summarise the statistics here and make comparison with the X-ray detections. The first Fermi-LAT catalogue (1FGL) (Abdo et al. 2010) includes 1451 $\gamma$-ray sources observed during the first 11 months of operation. Reducing this catalogue to the same region of sky covered by the AT20G survey and excluding known Galactic $\gamma$-ray sources leaves 540 extragalactic $\gamma$-ray sources in the southern sky. Of these, 233 (43\%) have an AT20G counterpart within the 95\% confidence ellipse. This is 4\% of the whole AT20G sample - very similar to the X-ray detection rate but only 56 (1\%) of the AT20G sources have both X-ray and $\gamma$-ray detections. The detection rate also increases with $\gamma$-ray flux density such that the brightest $\gamma$-ray sources are all detected in the AT20G survey. The median 20 GHz flux density for the $\gamma$-ray detected sources is 519 mJy, significantly brighter than the X-ray detections. Figure \ref{fig:alpha_X} shows the spectral index distribution of the Fermi-AT20G detections compared to the full AT20G sample. As expected, the sources that have gamma-ray counterparts are predominately flat-spectrum sources with median spectral indices $\alpha^{20}_{8}=-0.16$ and $\alpha^{8}_{5}=-0.01$. Kolmogorov-Smirnov tests confirm that these distributions are not drawn from the same population at the 99\% confidence level.

\begin{figure}
\begin{center}
\includegraphics[width=9cm]{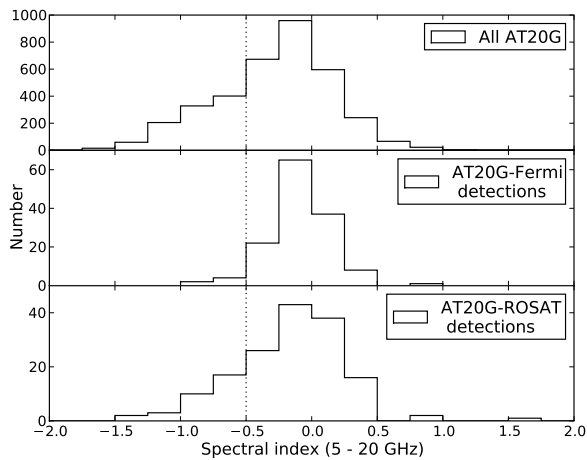}
\caption{5-20 GHz spectral index distribution for the full
AT20G sample (top panel), those sources with Fermi counterparts
(middle panel) and those with X-ray detections (bottom panel). The dotted line at $\alpha=-0.5$ denotes the commonly used dividing line to separate steep- and flat-spectrum sources.
} \label{fig:alpha_X}
\end{center}
\end{figure}

\section{Conclusions} \label{sec:Conclusions}
We analyzed a sample of 5808 extragalactic sources in the AT20G survey.

We present the source counts for the whole sample and for spectrally selected sub-samples. As expected, our analysis has confirmed that the high frequency bright source sample is dominated by flat-spectrum sources, while the fraction of steep-spectrum sources increases with decreasing flux density. As a consequence, the mean spectral index steepens and the spectral index distribution becomes more symmetric at lower flux density.
Extrapolations of the counts to different frequencies must consider flux density dependent steepening and changing shape of the spectral index distributions.

There is an overall spectral curvature causing spectral steepening of the population with frequency, which is more prominent at higher flux densities and at higher frequencies. The steepening is stronger for the more compact sources. The spectral distribution of the population of flat-spectrum sources is not a power law and changes with both flux density and redshift. Modelling this population will be a challenge.

The median fractional polarisation at 20 GHz is 2.7\% for $S_{20\ GHz}>250$ mJy. Above this flux density limit the sample is dominated by flat-spectrum sources. We find no evidence of a variation of the polarisation degree with flux density but, because of the very large number of upper limits, we could not extend this analysis to lower flux densities. Comparing the polarisation distributions for simultaneous detections at 5 and 20 GHz we find an indication of an average 17\% depolarisation at 5 GHz independent of flux density. Although spectral indices in total intensity and in polarisation are correlated, many, mainly flat-spectrum, sources have a spectral shape in polarisation substantially different from that in total intensity. However, sources with steep power law spectra in total intensity tend to preserve the same spectral shape in polarisation.

189 sources have a counterpart in the ROSAT-BSC catalogue and there is a general trend of increasing X-ray detection rate of AT20G sources with increasing 20 GHz flux density. There is no strong correlation between X-ray and radio flux densities. The distribution of radio spectral indices for the X-ray detected sources is similar to that for the total AT20G sample.

There are 233 AT20G sources that have gamma-ray counterparts in the first Fermi-LAT catalogue. The majority of these sources have flat spectral indices between 5 and 20\,GHz which is consistent with the high energy emission originating from compact cores of AGN.

Although both X-ray and $\gamma$-ray detection rates are related to the 20 GHz flux density, the $\gamma$-ray emission is more closely correlated with both the radio flux density and its spectrum. This suggests that the $\gamma$-ray and radio emission processes are more closely linked in space and time while the X-ray - radio link may be more indirect with AGN activity occuring in the same sources over a period of time.

\section*{Acknowledgments}
MM and GDZ acknowledge financial support for this research by ASI (ASI/INAF Agreement I/072/09/0 for the Planck LFI activity of Phase E2 and contract I/016/07/0 'COFIS').
TM acknowledges the support of an ARC Australian Postdoctoral Fellowship (DP0665973) and RDE acknowledges support of a Federation Fellowship (FF0345330).

We thank the staff at the Australia Telescope Compact Array site, Narrabri (NSW), for the valuable support they provide in running the telescope. The Australia Telescope Compact Array is part of the Australia Telescope which is funded by the Commonwealth of Australia for operation as a National Facility managed by CSIRO.

This research has made use of the NASA/IPAC Extragalactic Database (NED) which is operated by the Jet Propulsion Laboratory, California Institute of Technology, under contract with the National Aeronautics and Space Administration.

This research has made use of data obtained from the SuperCOSMOS Science Archive, prepared and hosted by the Wide Field Astronomy Unit, Institute for Astronomy, University of Edinburgh, which is funded by the UK Particle Physics and Astronomy Research Council.

\end{document}